\documentclass[jounal]{IEEEtran}
\usepackage{bm,amsmath,amssymb}
\usepackage{cite}
\ifCLASSINFOpdf
\usepackage{makecell}
\usepackage[pdftex]{graphicx}
\else
\usepackage[dvips]{graphicx}
\fi
\usepackage{subfigure}

\newtheorem{proposition}{Proposition}
\usepackage{xcolor}
\usepackage{colortbl,booktabs}
\usepackage{algorithm}
\usepackage{algorithmic}
\usepackage{courier}

\usepackage{courier}
\usepackage{multirow}
\usepackage{amssymb}
\usepackage{bbm,dsfont}
\usepackage{inconsolata}
\usepackage{stfloats}
\usepackage{epstopdf}

\begin{document}
	\title{Trainable Joint Channel Estimation, Detection and Decoding for MIMO URLLC Systems}
	\begin{sloppypar}
		\allowdisplaybreaks[4]
		%	\author{
			%		\IEEEauthorblockN{Yi Sun$^{1}$, Hong Shen$^{1}$, Wei Xu$^{1,2}$, and Chunming Zhao$^{1,2}$}
			%		\IEEEauthorblockA{$^1$ National Mobile Communications Research Laboratory, Southeast University, Nanjing, China}
			%		\IEEEauthorblockA{$^2$ Purple Mountain Laboratories, Nanjing, China}
			%		\IEEEauthorblockA{E-mail:\{sun\_yi, shhseu, wxu, cmzhao\}@seu.edu.cn}
			%	}
		
		\author{
			%Author 1, Author 2, and Author 3
			Yi~Sun,~\IEEEmembership{Student Member,~IEEE,} Hong~Shen,~\IEEEmembership{Member,~IEEE,} Bingqing~Li,~\IEEEmembership{Student Member,~IEEE,} Wei~Xu,~\IEEEmembership{Senior Member,~IEEE}, Pengcheng~Zhu,~\IEEEmembership{Member,~IEEE}, Nan~Hu, and~Chunming~Zhao,~\IEEEmembership{Member,~IEEE}
			% <-this % stops a space
			%\thanks{Author 1, Author 2, and Author 3 are with xxx (e-mail:\{Author 1, Author 2, Author 3\}@xxx).}
			
			\thanks{This work was supported in part by the National Key R\&D Program of China under Grant 2021YFB2900300, in part by the National Natural Science Foundation of China under Grant 62271137, and in part by the Southeast University China Mobile Research Institute Joint Innovation Center. ({\emph{Corresponding author: Hong Shen}}.)}
			
			\thanks{Y. Sun, H. Shen, B. Li, W. Xu, P. Zhu, and C. Zhao are with National Mobile Communications Research Laboratory, Southeast University, Nanjing 210096, China (e-mail:\{sun\_yi, shhseu, libingqing, wxu, p.zhu, cmzhao\}@seu.edu.cn). W. Xu, P. Zhu, and C. Zhao are also with Purple Mountain Laboratories, Nanjing 211111, China. N. Hu is with Institute of Wireless and Terminal Technology, China Mobile Research Institute, Beijing 100000, China (e-mail: hunan@chinamobile.com). }
			
			%		 <-this % stops a space	
		}

		\maketitle
		
		\begin{abstract}
			
			The receiver design for multi-input multi-output (MIMO) ultra-reliable and low-latency communication (URLLC)  systems can be a tough task due to the use of short channel codes and few pilot symbols. Consequently, error propagation can occur in traditional turbo receivers, leading to performance degradation. { Moreover, the processing delay induced by information exchange between different modules may also be undesirable for URLLC.} To address the issues, we advocate to perform joint channel estimation, detection, and decoding (JCDD) for MIMO URLLC systems encoded by short low-density parity-check (LDPC) codes. Specifically, we develop two novel JCDD problem formulations based on the maximum \emph{a posteriori} (MAP) criterion for Gaussian MIMO channels and sparse mmWave MIMO channels, respectively, which integrate the pilots, the bit-to-symbol mapping, the LDPC code constraints, as well as the channel statistical information. Both the challenging large-scale non-convex problems are then solved based on the alternating direction method of multipliers (ADMM) algorithms, where closed-form solutions are achieved in each ADMM iteration. Furthermore, two JCDD neural networks, called JCDDNet-G and JCDDNet-S, are built by unfolding the derived ADMM algorithms and introducing trainable parameters. It is interesting to find via simulations that the proposed trainable JCDD receivers can outperform the turbo receivers with affordable computational complexities.
		\end{abstract}
		
		
		\section{Introduction}
		Ultra-reliable low-latency communications (URLLC), as a key enabler of the fifth generation (5G) and beyond 5G systems, are expected to support some mission-critical applications with stringent requirements, such as remote surgery, intelligent transport, and factory automation \cite{5G8403963,URLLC,URLLC789}. For example, the reliability is required to be no smaller than 99.999\% with the end-to-end latency limited to within 1 millisecond (ms) \cite{5G8403963}. To achieve high reliability of URLLC, multi-input multi-output (MIMO) techniques can be a promising solution by fully exploiting %equipping multiple receive antennas and transmit antennas, which can exploit 
		the spatial diversity gain and the antenna array gain. On the other hand, short packet transmission can be adopted % to achieve the desired high reliability and low latency, respectively. Moreover, another efficient way 
		to meet the latency requirement of URLLC, which has also been extensively studied in literature \cite{URLLCSPC,SPC8944280,SPC1275673}. Taking advantages of both techniques, short packet MIMO transmission has emerged as an attractive topic in the context of URLLC.

\begin{figure}[!t]
			
			\centering
			\includegraphics[width=\linewidth]{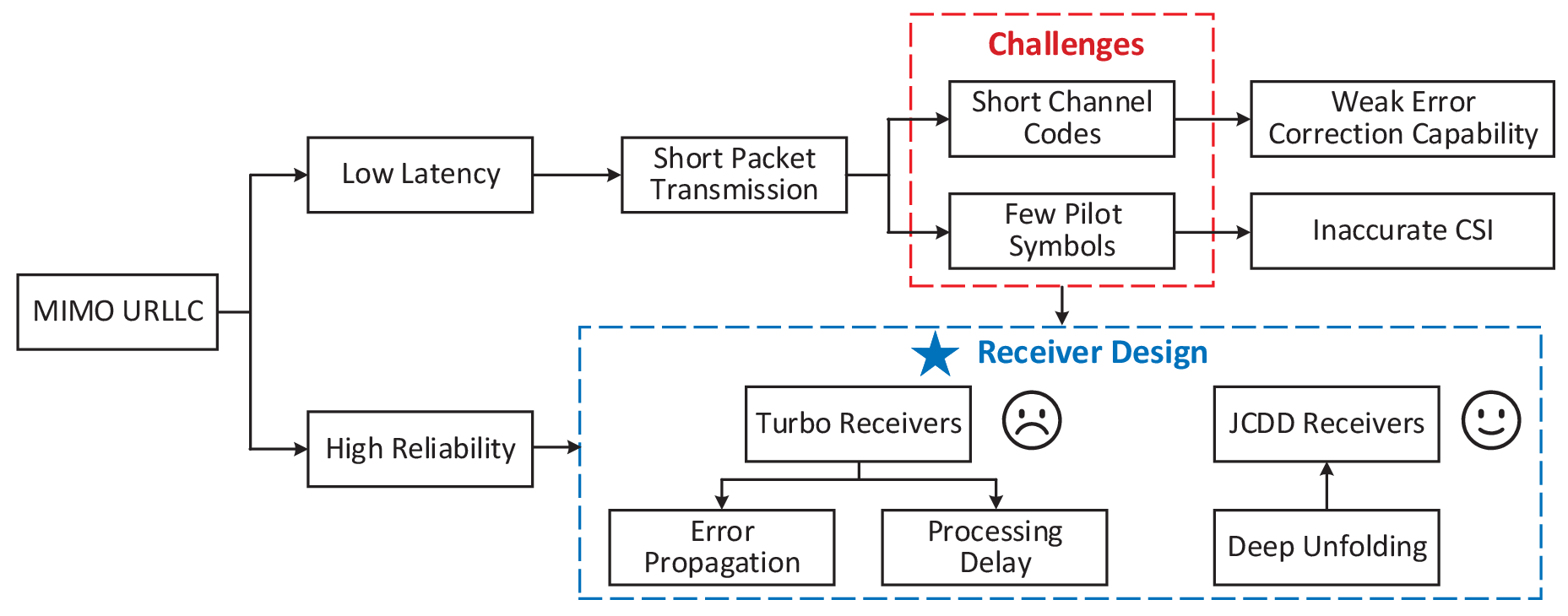}
			\caption{The motivation of the proposed receiver design.}
			\label{fig_all}
		\end{figure}

		The classical information-theoretical metrics, such as the ergodic capacity and the outage capacity, are established based on the Shannon theory under the assumption of infinite blocklength, which are no longer suitable for URLLC systems with short packets. %Conventional communication systems are designed based on the Shannon theory under the assumption of infinite blocklength, which, obviously, does not hold for short packets. Therefore, some classical information-theoretical metrics, such as the ergodic capacity and the outage capacity, are no longer suitable for MIMO URLLC systems. 
		To overcome the limitation, a new performance metric, namely the maximal channel coding rate with given blocklength and error probability, was introduced in \cite{FINITE8640815}. Followed by this, the performance analysis for short packet transmissions over quasi-static MIMO fading channels was studied in \cite{quasi2014,SPC8944280}, and the extension to massive MIMO systems was further presented in \cite{URLLC111}. %, where the impact of imperfect channel state information (CSI), pilot contamination, spatially correlated channels, and arbitrary linear spatial processing were incorporated. % for the characterization of the error probability achievable at a finite blocklength. 
		Although these works have ascertained the theoretical limits for MIMO URLLC systems, the receiver designs approaching such limits have not been thoroughly investigated. %A linear minimum error probability (MEP) detector with finite blocklength and imperfect CSI and a transceiver for full-duplex (FD) URLLC systems were derived in \cite{linear2019} and \cite{transceiver2022}, respectively. Nevertheless, both of them were developed by optimizing the information-theoretical metrics, while the performance gains in terms of more practical and intuitive metrics, such as the block error rate (BLER), have not been clarified. Nevertheless, the channel codes for error correction, as an essential part in a practical system, were not taken into account therein. 
		{{In fact, only short channel codes are allowed to match the short packet transmission, which can cause performance deterioration due to their relatively weak error correction capabilities \cite{ShortCode}. On the other hand, the reliability of a receiver relies on the quality of channel state information (CSI), which is generally acquired based on the pilot-aided channel estimation. However, the number of pilots should be small to guarantee a low transmission delay, leading to inaccurate CSI. Therefore,  how to design an efficient MIMO URLLC receiver in the presence of short codes and few pilots remains a critical and challenging issue. To be clear, Fig. \ref{fig_all} provides the motivation of our work.}}
		
		%		\begin{figure}[t]
			%			{\centering
				%				\includegraphics[width=0.4\textwidth]{frame.eps}
				%				\caption{Frame structures of (a) traditional long packet transmission and (b) short packet transmission.}
				%				\label{fig0}}
			%		\end{figure}
		
\begin{figure}[!t]
			
			\centering
			\subfigure[Traditional turbo receivers]
			{\begin{minipage}[t]{0.5\textwidth}
					\centering
					\includegraphics[width=8cm]{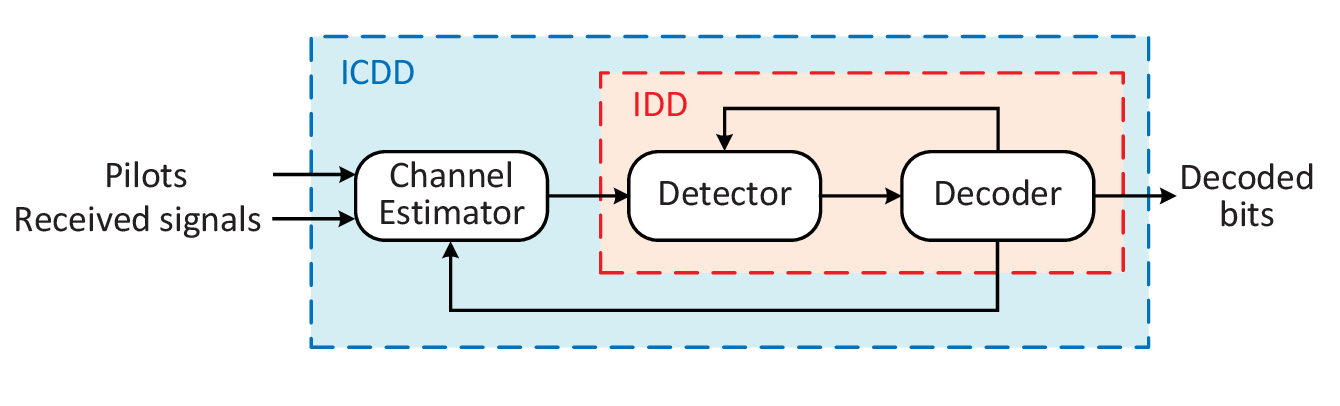}
				\end{minipage}		
			}\quad
			\subfigure[Proposed JCDD receivers]
			{\begin{minipage}[t]{0.5\textwidth}
					\centering
					\includegraphics[width=8cm]{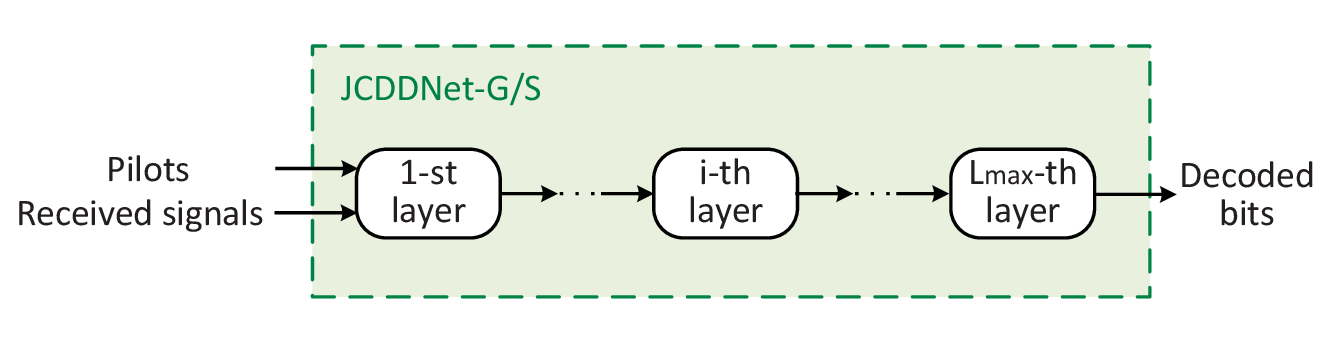}
				\end{minipage}		
			}	
			\caption{Block diagrams of (a) traditional turbo receivers and (b) proposed trainable JCDD receivers.}
			\label{fig1}
		\end{figure}

		In traditional MIMO receivers, the channel estimation, the signal detection, and the channel decoding are performed in a decoupled and sequential manner. Although the three modules have been extensively studied in previous literature \cite{MIMO,Coding}, %we can easily find an efficient algorithm for each of them. However, 
		the overall receiver performance can be far from the optimum due to the disjoint nature. Alternatively, iterative detection and decoding (IDD), also known as the turbo receiver, was proposed to achieve near-optimal performance \cite{Turbo2002,ASIC7894280,Achieving2003}. {As depicted in the red block in Fig. \ref{fig1}(a),} in each turbo iteration, the extrinsic soft information is exchanged between the detector and the decoder to enhance the performance. Nevertheless, the inaccurate channel estimates caused by the limited number of pilot symbols can result in an unreliable information exchange in IDD and, furthermore, performance degradation. To handle this, a more generalized turbo receiver was developed by including the channel estimation into the loop \cite{MICED}, {which is also shown in the blue block in Fig. \ref{fig1}(a).} 
		Specifically, the feedback of the decoder is available to both the detector and the channel estimator to enable the iterative channel estimation, detection, and decoding (ICDD), so as to refine the CSI quality as well as the subsequent detection and decoding performance. 
		
		Despite the empirical successes of turbo receivers under long packet transmissions, the performance can deteriorate in MIMO URLLC systems encoded by short low-density parity-check (LDPC) codes, where the belief propagation (BP) decoder may suffer from undesirable positive feedback due to the short cycles in the Tanner graph, leading to severe error propagation. In order to mitigate the impacts of decoding errors during turbo iterations, a mean squared error (MSE) based virtual pilot selection strategy was designed in \cite{SPCreceiver} to utilize part of the most reliable data symbols for iterative channel estimation. However, non-negligible processing delay can be introduced by the iterative information exchange between receiver modules, which does not conform to the requirement of URLLC. To address these concerns, a new type of receivers were developed for coded MIMO systems by integrating the detection and the decoding in a  %formulating 
		linear programming (LP) formulation %, where the binary code constraints were transformed into linear inequalities in real field to facilitate the integration of detection and decoding 
		\cite{LDPC13,Polar86}. Furthermore, the authors extended the studies by additionally incorporating the pilot symbol constraints and the noise subspace constraints in \cite{LDPC8715338}. However, the $\mathcal{L}_1$-norm based objective function and constraints used in these works to adapt to the LP framework are suboptimal in terms of the receiver performance. Besides, these LP receivers entail high computational complexity, which thus hinders their practicality. {On the other hand, inspired by the great potential of deep learning-based communications \cite{BELL1,BELL2}, a fully convolutional neural network (CNN)-based receiver DeepRx was proposed for orthogonal frequency division multiplexing (OFDM) systems\cite{DeepRx}. Furthermore, the authors of \cite{MLRx} also leveraged CNN to enhance the traditional MIMO-OFDM receivers. Similar ideas were studied in \cite{DLRx} to combat the power amplifier-induced nonlinear distortion. Nevertheless, these learning-aided receivers are unable to exploit the decoding information due to the absence of the feedback links. An end-to-end autoencoder was developed in \cite{JDED} for joint detection, equalization and decoding in short-packet transmissions. Albeit with considerable performance improvement, the work can hardly be extended to MIMO systems due to the excessively large network scale. Additionally, the coding and modulation at the transmitter were totally replaced by data-driven structures in the autoencoder, rendering it incompatible with the existing communication systems. } 
		
		In this paper, we propose two trainable joint channel estimation, detection, and decoding (JCDD) receivers for Gaussian MIMO channels and  sparse mmWave MIMO channels, respectively, which are fundamentally distinct from previous efforts in the following aspects. First, the proposed receivers integrate the functions of different modules, which can effectively avoid the error propagation and the processing delay in the turbo receivers. Second, a unified maximum \emph{a posteriori} (MAP)-based formulation is considered, which can promise a better receiver performance than the LP-based formulation with the $\mathcal{L}_1$-norm metrics. Third, instead of the end-to-end structure in a fully data-driven manner, we only focus on the receiver design and build a model-driven JCDD network based on the popular deep unfolding technique \cite{Model5338,unroll}, which can inherit the expert knowledge and also relieve the training burden. Our main contributions include: 
		\begin{itemize}
			\item Following the MAP criterion and embracing the channel statistical information as well as the constraints imposed on pilots, data symbols, and LDPC codes, two JCDD problems are formulated for Gaussian MIMO channels and sparse mmWave MIMO channels, respectively, which, to the best of our knowledge, have not been investigated in previous works.
			\item Motivated by the wide applications of the alternating direction method of multipliers (ADMM) in solving large-scale optimization problems \cite{ADMMBook}, especially for LDPC decoding \cite{ADMMDecoder}, we derive two ADMM-based algorithms to solve the considered JCDD problems, which only requires the calculations of closed-form expressions in each ADMM iteration.
			\item {By unfolding the ADMM iterations, two model-driven JCDD networks, i.e., JCDDNet-G and JCDDNet-S, are constructed as shown in Fig. \ref{fig1}(b), where the involved parameters can be optimized via offline training.} Furthermore, a relaxed and accelerated ADMM (R-A-ADMM) structure is utilized to improve the ADMM convergence. Besides, we adopt a multi-stage multi-layer training strategy to strike a tradeoff between the training complexity and the overall performance, which is also friendly to the early-termination mechanism.
		\end{itemize}

		The rest of this paper is organized as follows. Section II describes the system model. In Section III and Section IV, we establish two JCDD receivers for Gaussian MIMO channels and sparse mmWave MIMO channels, respectively, wherein their corresponding deep unfolding networks are also developed. The extension for the multiuser case is studied in Section V. Section VI presents the numerical results. Finally, we conclude this work in Section VII. 
		
		\section{System Model}
		{Consider a coded single-carrier MIMO system equipped with $N_t$ transmit antennas and $N_r$ receive antennas, which can also be regarded as an OFDM system where the channel is relatively flat in time and frequency domains as in \cite{MICED}.} The channel coefficients are assumed to be constant within a transmission block that occupies $T={T_P}+{T_D}$ time slots, where $T_P$ and $T_D$ time slots are assigned for the transmission of pilot and data symbols, respectively. For the data transmission, we first encode $K$ information bits by a binary encoder with code length $N$. Specifically, in this paper, we adopt the widely-used LDPC code for its excellent error correction performance. Let ${\bf{H}} \in {\left[0,1\right]}^{M \times N}$ be the corresponding binary parity-check matrix of an LDPC code, where $M$ is the number of parity-check nodes. Then, a valid codeword ${\bf{b}} = {[{b_1},{b_2}, \cdots ,{b_N}]^T}$ can be specified by
		\begin{equation}
			\left[\sum\limits_{i = 1}^N {{H_{ji}}{b_i}}\right]_2 = 0,\quad j = 1,2, \cdots ,M,\label{eq1}
		\end{equation}
		%\vspace{-1em}
		\begin{equation}
			{\bf{b}} \in \{ 0,1\}^N,\label{eq2}
		\end{equation}
		where $[\cdot]_2$ denotes the modular-2 operator. After the LDPC encoding is performed, every $Q$ coded bits are mapped into a $2^Q$-quadrature amplitude modulation (QAM) symbol, resulting in a transmit vector ${\bf{x}} = {[{x_1},{x_2}, \cdots ,{x_{N/Q}}]^T}$. Finally, these modulated symbols are demultiplexed into $N_t$ data streams of length ${T_D} = N/{N_t}/Q$  and sent over the MIMO channel ${{\bf G}} \in \mathbb{C}^{{N_r}\times{N_t}}$. Denote the pilot matrix and the data matrix by ${{\bf S}_P} \in \mathbb{C}^{{N_t}\times{T_P}}$ and ${{\bf S}_D} \in \mathbb{C}^{{N_t}\times{T_D}}$, respectively. Then, the received signals can be expressed by
		\begin{equation} 
			\left[ {{{\bf{Y}}_P},{{\bf{Y}}_D}} \right] = {\bf{G}}\left[ {{\bf{S}}_P,{\bf{S}}_D} \right] + {\bf{N}},\label{eq3}
		\end{equation}
		where ${{\bf{Y}}_P}  \in \mathbb{C}^{{N_r}\times{T_P}}$ and ${{\bf{Y}}_D}  \in \mathbb{C}^{{N_r}\times{T_D}}$ are the received pilot matrix and data matrix, respectively, and ${\bf{N}}  \in \mathbb{C}^{{N_r}\times{T}}$ is the additive white Gaussian noise (AWGN) matrix whose elements follow the independently identically distributed (i.i.d.) Gaussian distribution with zero mean and variance ${\sigma}^2$. Note that the $(k,t)$-th entry of ${\bf S}_D$, i.e., $s^D_{kt}$, corresponds to the QAM symbol $x_{{i^{t,k}}}$, where ${i^{t,k}} \triangleq {N_t}(t - 1) + k$. 
		
		The ultimate goal of the receiver is to recover the information bits based on the received signals in (\ref{eq3}) with a given pilot matrix $\bf P$, which is challenging since both $\bf G$ and $\bf D$ are unknown to the receiver. As mentioned in the introduction part, although turbo receivers can boost the performance via information exchange between different modules, they suffer from error propagation especially for short codes with relatively low error correction capabilities. To overcome this shortcoming, we %propose to perform JCDD based on the MAP formulations, which can simultaneously leverage the channel statistical information and the constraints imposed on pilots, data symbols, and LDPC codes. 
		establish a MAP based JCDD receiver framework that unifies the channel estimation, the data detection, and the LDPC decoding. The details of the proposed receiver designs for Gaussian MIMO channels and mmWave MIMO channels are elaborated in the following two sections, respectively.

		\section{JCDD for Gaussian MIMO Channels}
		A JCDD receiver for Gaussian MIMO channels is developed in this section. We first formulate a JCDD problem using the MAP criterion and then derive an ADMM-based solution to solve this difficult mixed-integer problem. Based on this, we further present a model-driven network JCDDNet-G to facilitate the optimization of the involved parameters.
		
		\subsection{Problem Formulation}
		With the received signals ${{\bf{Y}}_P}$ and ${{\bf{Y}}_D}$ available, we first formulate a MAP based joint channel estimation and detection (JCD) problem as follows: 
		\begin{equation} 
			%		\begin{small}
				\begin{aligned}
					\left[ {\bf{G}},{\bf{S}}_D \right]=&\mathop {\arg \max }\limits_{{\bf{G}},{\bf{S}}_D \in {\Omega ^{{N_t} \times {T_D}}}} p\left( {{\bf{G}},{\bf{S}}_D\left| {{{\bf{Y}}_P},{{\bf{Y}}_D},{\bf{S}}_P} \right.} \right)\\
					\mathop  = \limits^{(a)} &\mathop {\arg \max }\limits_{{\bf{G}},{\bf{S}}_D \in {\Omega ^{{N_t} \times {T_D}}}} p\left( {\left[ {{{\bf{Y}}_P},{{\bf{Y}}_D}} \right]\left| {{\bf{G}},\left[ {{\bf{S}}_P,{\bf{S}}_D} \right]} \right.} \right)p\left( {\bf{G}} \right),\label{eq4}
				\end{aligned}
				%\end{small}
			\end{equation}
			where (a) is based on the Bayes' theorem and the assumption that each entry of ${\bf{S}}_D$ is equally drawn from a discrete QAM constellation set $\Omega$. Letting ${\bf Y} \triangleq {\left[ {{{\bf{Y}}_P},{{\bf{Y}}_D}} \right]}$ and ${{\bf S}} \triangleq {\left[ {{\bf{S}}_P,{\bf{S}}_D} \right]}$, the likelihood probability in (\ref{eq4}) is given by
			\begin{equation} 
				\begin{small}
					\begin{aligned}
						%p\left( {\left[ {{{\bf{Y}}_P},{{\bf{Y}}_D}} \right]\left| {{\bf{G}},\left[ {{\bf{P}},{\bf{D}}} \right]} \right.} \right)
						p\left( {{\bf{Y}}\left| {{\bf{G}},{\bf S}} \right.} \right)
						=\frac{1}{{{\left(\pi{\sigma^2}\right) ^{{N_r}T}}}}\exp \left(-\frac{\left\| {{{\bf y}} - \left({{\bf S}^T}\otimes{{\bf I}_{N_r}}\right){\bf{g}}}\right\|_2^2}{\sigma ^2} \right),\label{eq5}
					\end{aligned}
				\end{small}
			\end{equation}
			where ${\bf y} \triangleq {\left[ {{{\bf{y}}^T_P},{{\bf{y}}^T_D}} \right]^T}$, ${{\bf y}_P}\in \mathbb{C}^{{N_r}{T_P} \times 1}$, ${{\bf y}_D}\in \mathbb{C}^{{N_r}{T_D} \times 1}$, and ${\bf g}\in \mathbb{C}^{{N_r}{N_t} \times 1}$ are the vectorized form of ${\bf Y}_P$, ${\bf Y}_D$, and $\bf G$, respectively, %${\left\| \cdot \right\|_F}$ denotes the Frobenius norm of the input matrix, 
			and ${\left\| \cdot \right\|_2}$ denotes the $\mathcal{L}_2$ norm of the input vector. Assume that $\bf g$ follows a zero-mean Gaussian distribution with covariance ${{{\bf{C}}_{\bf{g}}}}$. Then, we have
			\begin{equation} 
				%	\begin{small}
					\begin{aligned}
						p\left( {\bf{g}} \right) = \frac{1}{{{\pi ^{{N_r}{N_t}}}\left| {{{\bf{C}}_{\bf{g}}}} \right|}}{e^{ - {{\bf{g}}^H}{\bf{C}}_{\bf{g}}^{ - 1}{\bf{g}}}}.\label{eq6}
					\end{aligned}
					%\end{small}
				\end{equation}
				By substituting (\ref{eq5}) and (\ref{eq6}) into (\ref{eq4}) and performing some mathematical manipulations, we obtain the following MAP-JCD problem:
				\begin{equation}
					%			\begin{small}
						\begin{aligned}
							\mathop {\min }\limits_{{\bf{g}},{\bf{S}}_D \in {\Omega ^{{N_t} \times {T_D}}}}{{\left\| {{\bf y}} - \left({{\bf S}^T}\otimes{{\bf I}_{N_r}}\right){\bf{g}} \right\|_2^2}+{\sigma ^2}{{\bf{g}}^H}{\bf{C}}_{\bf{g}}^{ - 1}{\bf{g}}}.
							\label{eq7} 
						\end{aligned}
						%\end{small}
					\end{equation}
					%as ${\cal{CN}}\left({\bf 0},{\bf C}_{\bf g}\right)$.
					
					To further achieve JCDD, we need to formulate a bit-level receiver by incorporating the code constraints into (\ref{eq7}). Specifically, we first introduce a bit-to-symbol mapping function ${\bf S}_D = f({\bf b})$, which can be element-wisely expressed as
					\begin{equation} 
						%		\begin{small}
							\begin{aligned}
								s^D_{kt} = f({{\bf{\tilde b}}^{t,k}}), \quad k=1,2,\cdots,N_t,\quad t=1,2,\cdots,{T_D},\label{eq8}
							\end{aligned}
							%\end{small}
						\end{equation}
						where ${{\bf{\tilde b}}^{t,k}} = [{b_{Q({{{i^{t,k}}}}-1) + 1}}, {b_{Q({{{i^{t,k}}}}-1) + 2}}, \cdots ,{b_{Q{{{i^{t,k}}}}}}]\in {\{ 0,1\} ^Q}$ with ${i^{t,k}} \triangleq {N_t}(t - 1) + k$. %The Gray mapping is considered in this paper. For example, 
						{ We adopt the Gray mapping functions according to \cite{3GPP}, which are given in Appendix \ref{Ap_1}.} 
						
						By plugging  (\ref{eq8}) into the objective function of (\ref{eq7}) and including the LDPC code constraints, we achieve the following MAP-JCDD formulation:
						\begin{equation} 
							\begin{aligned}
								\mathop {\min }\limits_{{\bf{g}},{\bf b}}\quad&{{\left\| {{\bf y}} - \left({\left[ {{\bf{S}}_P,f({\bf b})} \right]^T}\otimes{{\bf I}_{N_r}}\right){\bf{g}} \right\|_2^2}+{\sigma ^2}{{\bf{g}}^H}{\bf{C}}_{\bf{g}}^{ - 1}{\bf{g}}}
								\\{\rm{s.t.}} \quad &(1),(2). \label{eq11}
							\end{aligned}
						\end{equation}
						
						Unfortunately, it is difficult to solve problem (\ref{eq11}). First, the objective function is non-convex due to the coupled variables. Second, the optimization variables are of high dimensions. Third, the parity check constraints in (\ref{eq1}) and the discrete constraints in (\ref{eq2}) are hard to handle. In the following subsection, we provide an efficient ADMM-based solution to this problem.
						
						\subsection{ADMM-Based Solution}
						To make problem (\ref{eq11}) more tractable, we first relax the LDPC code constraints using the corresponding fundamental parity polytopes, which can be characterized by the following inequalities \cite{LPcodes}{\footnote{ The well-known polar codes with dense parity check matrices can also be used by following the idea in \cite{LPpolar}, which utilizes the recursive nature of polar codes to obtain a sparse graph representation with ${\cal O}\left(N \log N\right)$ auxiliary variables. Based on the graph, new polytopes with a relatively small number of constraints can be defined.}} 
						\begin{equation} 
							\begin{aligned}
								\sum\limits_{i \in {{{\cal F}}(j)}} {{b_i}}  -& \sum\limits_{i \in {{{\cal N}_c}(j)}\backslash {{\cal F}}(j)} {{b_i}}  \le \left| {{{\cal F}}(j)} \right| - 1,\\
								&\forall {{{\cal F}}(j)}\subseteq {\cal S}(j), \quad j = 1,2, \cdots ,M,\label{eq12}
							\end{aligned}
						\end{equation}
						and the boxed constraints
						\begin{equation}
							{\bf b} \in [0,1]^N, \label{eq13}
						\end{equation}
						where ${{\cal N}_c}(j)$ stands for the set of bit variables involved in the $j$-th parity check constraint, and ${\cal S}(j)$ is defined as the union set of all ${{\cal N}_c}(j)$'s subsets with odd cardinalities. It can be inferred that the number of inequalities corresponding to the $j$-th parity check constraint is ${{2^{\left| {{{\cal N}_c}(j)} \right| - 1}}}$. For ease of the subsequent derivation, we further rewrite (\ref{eq12}) into a matrix form as
						\begin{equation} 
							{\bf A}{\bf b} \leq {\bm \theta},\label{eq14}
						\end{equation}
						where ${\bf A} = \left[[{{{\bf W}_1}{{\bf Q}_1}}]^T,[{{{\bf W}_2}{{\bf Q}_2}}]^T,\cdots,[{{{\bf W}_M}{{\bf Q}_M}}]^T\right]^T$ and ${\bm \theta} = \left[{{{\bm \theta}^T_1}},{{{\bm \theta}^T_2}},\cdots,{{{\bm \theta}^T_M}}\right]^T$, with ${{\bf W}_j} \in {{\left\{-1,1\right\}}^{{{2^{\left| {{{\cal N}_c}(j)} \right| - 1}}} \times {\left| {{{\cal N}_c}(j)} \right|}}}$, ${{\bf Q}_j}\in {{\left\{0,1\right\}}^{{\left| {{{\cal N}_c}(j)} \right|} \times N}}$, and ${{\bm \theta}_j} \in {\mathbb R}^{ {{2^{\left| {{{\cal N}_c}(j)} \right| - 1}}}}$ being the weight matrix, the variable-selection matrix used for extracting the corresponding bit variables, and the bias matrix related to the $j$-th parity check constraint, respectively.\footnote{To be clear, we take a 3-variable parity check constraint for example. Assume that the $j$-th parity check constraint is ${\left[{b_2}+{b_{10}}+{b_{36} }\right]_2}=0,{b_2},{b_{10}},{b_{36}}\in{\left\{0,1\right\}}$ . Then the corresponding ${{\bf W}_j}= \left[1,-1,-1;-1,1,-1;-1,-1,1;1,1,1\right]$, %\begin{bmatrix} 1 & -1 & -1 \\ -1 & 1 & -1 \\ -1 & -1 & 1 \\ 1 & 1 & 1 \end{bmatrix}
							${{\bf Q}_j}$ is the all-zeros matrix of size $3 \times N$ except the $(1,2)$-th, $(2,10)$-th and $(3,36)$-th entries being $1$, and ${{\bm \theta}_j} = \left[0,0,0,2\right]^T$.} 
						Here we abuse the use of the notation ``$\leq$'' to represent that each element of the left-side vector is no larger than that of the right-side vector. Moreover, we also mention that $\bf A$ is sparse and column-wise orthogonal, whose elements are either $0$, $-1$ or $1$ \cite{3var_LDPC}.
						
						Furthermore, we additionally impose a quadratic penalty term $-\alpha \left\| {{\bf{b}} - {\bf{0.5}}_N} \right\|_2^2$ to the objective function of (\ref{eq11}) to encourage the integer solution of each element of $\bf b$ and therefore tighten the boxed constraints in (\ref{eq13}), where $\alpha>0$ is the corresponding penalty parameter and ${\bf{0.5}}_N$ represents the vector of length $N$ with each element being $0.5$. Consequently, we are able to reformulate problem (\ref{eq11}) by
						\begin{equation} 
							%	\begin{small}
								\begin{aligned}
									\mathop {\min }\limits_{{\bf{g}},{\bf b}}\quad&{{\left\| {{\bf y}} - {{\bf X}_{\bf b}}{\bf{g}} \right\|_2^2}+{\sigma ^2}{{\bf{g}}^H}{\bf{C}}_{\bf{g}}^{ - 1}{\bf{g}}}-\alpha \left\| {{\bf{b}} - {\bf{0.5}}_N} \right\|_2^2
									\\\rm{s.t.}\quad&(13),(14), \label{eq15}
								\end{aligned}
								%\end{small}
							\end{equation}
							where ${{\bf X}_{\bf b}} \triangleq {{\bf S}_{\bf b}^T}\otimes{{\bf I}_{N_r}}$, ${\bf S}_{\bf b} \triangleq \left[ {{\bf{S}}_P,f({\bf b})} \right]$. { Thanks to the continuous variables and the introduced quadratic penalty term, the reformulation gets easier to handle and can be equivalent to problem (\ref{eq11}) for a sufficiently large $\alpha$. However, due to the non-convexity of $-\alpha \left\| {{\bf{b}} - {\bf{0.5}}_N} \right\|_2^2$, a large $\alpha$ might yield a poor local minimum, while a small $\alpha$ leads to a less accurate approximation of the original binary problem \cite{Binary}. Hence, the value of $\alpha$ should be carefully selected.}
							
							Note that problem (\ref{eq15}) is convex quadratic with respect to (w.r.t.) $\bf g$. Thus, the optimal solution to $\bf g$ can be obtained as
							\begin{equation} 
								{\hat{\bf g}} = {\left( {{{\bf{X}}_{\bf b}^H}{\bf{X}}_{\bf b} + {\sigma ^2}{\bf{C}}_{\bf{g}}^{ - 1}} \right)^{ - 1}}{{\bf{X}}_{\bf b}^H}{\bf{y}}.\label{eq16}
							\end{equation}
							Then, by substituting (\ref{eq16}) back to the objective function in (\ref{eq15}) and safely dropping the constant term, problem (\ref{eq15}) can be equivalently simplified to a problem dependent on $\bf b$ as
							\begin{equation} 
								\begin{small}
									\begin{aligned}
										\mathop {\min}\limits_{\bf b}\quad&-{{{\bf{y}}^H}{{\bf{X}}_{\bf b}}{{\left( {{{\bf{X}}_{\bf b}^H}{{\bf{X}}_{\bf b}} + {\sigma ^2}{\bf{C}}_{\bf{g}}^{ - 1}} \right)}^{ - 1}}{{\bf{X}}_{\bf b}^H}{\bf{y}}}-\alpha \left\| {{\bf{b}} - {\bf{0.5}}_N} \right\|_2^2
										\\\rm{s.t.}\quad&(13),(14). \label{eq17}
									\end{aligned}
								\end{small}
							\end{equation}
							%	which is a large-scale optimization problem dependent on $\bf b$.
							
							\newcounter{TempEqCnt}
							\setcounter{TempEqCnt}{\value{equation}} 
							\setcounter{equation}{19} 
							\begin{figure*}[ht]
								\begin{equation}
									\begin{small}
										\begin{aligned}		
											{\left({{{\tilde{\bf b}}}^{t,k}}\right)^n} = \mathop {\arg \min }\limits_{{{{\bf{\tilde b}}}^{t,k}} \in {{[0,1]}^Q}} {\lambda ^{n - 1}}{\left| {f({{{\bf{\tilde b}}}^{t,k}})} \right|^2} &- 2{\rm{Re}}\left\{ {{{\left( {f({{{\bf{\tilde b}}}^{t,k}})} \right)}^*}s_{kt}^{n - 1}} \right\} - \alpha \sum\limits_{q = 1}^Q {\left( {{b_{{i^{t,k,q}}}^2} - {b_{{i^{t,k,q}}}}} \right)} \\ & + \frac{\mu }{2}\sum\limits_{q = 1}^Q {\left( {\Lambda_{i^{t,k,q}}}{b_{{i^{t,k,q}}}^2 + 2{\bf{a}}_{{i^{t,k,q}}}^T\left( {{{\bf{z}}^{n - 1}} - {\bm{\theta }} + {{\bm{\eta }}^{n - 1}}} \right){b_{{i^{t,k,q}}}}} \right)},%\\ &\quad\quad\quad\quad\quad\quad\quad\quad\quad\quad k=1,2,\cdots,N_t,\quad t=1,2,\cdots,{T_D},
											\label{eq26}
										\end{aligned}
									\end{small}
								\end{equation}
								\hrulefill
							\end{figure*}
							\setcounter{equation}{\value{TempEqCnt}} 
							
							Now we develop an ADMM-based solution to problem (\ref{eq17}). First, we introduce an auxiliary variable ${\bf z}\in {\mathbb R}^{\Gamma^c}$ with $\Gamma^c = { \sum_{j=1}^M{{2^{\left| {{{\cal N}_c}(j)} \right| - 1}}}}$ and recast problem (\ref{eq17}) by
							\begin{equation} 
								\begin{aligned}
									\mathop {\min }\limits_{{\bf{b}} \in {\left[0,1\right] ^{{N} \times 1}}}\quad&\Phi({\bf b})-\alpha \left\| {{\bf{b}} - {\bf{0.5}}_N} \right\|_2^2
									\\\text{s.t.}\quad\quad&{\bf A}{\bf b}+ {\bf z}= {\bm \theta},\quad{\bf z}\geq{\bf 0}, \label{eq18}
								\end{aligned}
							\end{equation}
							where $\Phi({\bf b})=-{{{\bf{y}}^H}{{\bf{X}}_{\bf b}}{{\left( {{{\bf{X}}_{\bf b}^H}{{\bf{X}}_{\bf b}} + {\sigma ^2}{\bf{C}}_{\bf{g}}^{ - 1}} \right)}^{ - 1}}{{\bf{X}}_{\bf b}^H}{\bf{y}}}$. Based on this representation, the corresponding scaled augmented Lagrangian function can be constructed as 
							\begin{equation} 
								\begin{aligned}
									L_{\mu}\left({\bf{b}},{\bf{z}},{\bm{\eta}}\right) = \Phi({\bf b})& - \alpha \left\| {{\bf{b}} - {{\bf{0.5}}_N}} \right\|_2^2 + \\
									&\frac{\mu }{2}{\left\| {\bf A}{\bf b}+ {\bf z} - {\bm \theta} + {\bm \eta} \right\|_2^2}- \frac{\mu }{2} {\left\| {\bm \eta}\right\|_2^2}, \label{eq19}
								\end{aligned}
							\end{equation}
							where ${\bm{\eta}} \in {\mathbb{R}}^{\Gamma^c}$ and ${\mu}>0$ denote the scaled dual variables and the penalty parameter, respectively. Accordingly, the $n$-th ADMM iteration can be described as follows \cite{ADMMBook}:
							\begin{subequations}
								\label{eq20}
								\begin{equation}
									%				\begin{small}
										\begin{aligned}
											{{\bf{b}}^{n}} = \mathop {\arg \min }\limits_{{\bf{b}} \in {{[0,1]}^N}} \Phi({\bf b}) &- \alpha \left\| {{\bf{b}} - {{\bf{0.5}}_N}} \right\|_2^2 \\
											&+ \frac{\mu }{2}{\left\| {\bf A}{\bf b}+ {{\bf{z}}^{n-1}} - {\bm \theta} + {{\bm \eta}^{n-1}} \right\|_2^2}, \label{eq20a}
										\end{aligned}
										%			\end{small}	
								\end{equation}
								\begin{equation}
									%				\begin{small}
										\begin{aligned}
											{\bf{z}}^{n} = {\prod\nolimits_{{\left[0,\infty\right)}^{\Gamma^c}}}\left({\bm \theta} - {\bf A}{{\bf b}^{n}}- {\bm{\eta}}^{n-1}\right),\label{eq20b}
										\end{aligned}
										%		\end{small}	
								\end{equation}
								\begin{equation}
									%				\begin{small}
										\begin{aligned}
											{\bm{\eta }}^{n} = {\bm{\eta }}^{n-1} + {\bf A}{{\bf b}^{n}}+ {{\bf z}^{n}} - {\bm \theta},\label{eq20c}
										\end{aligned}
										%		\end{small}	
								\end{equation}
							\end{subequations}
							where ${\prod\nolimits_{{\left[0,\infty\right)}^{\Gamma^c}}}(\cdot)$ represents projecting each element of the input vector onto the interval $\left[0,\infty\right)$. 	
							
							The remaining task is to tackle subproblem (\ref{eq20a}), where the challenge mainly lies in the non-trivial term $\Phi({\bf b})=-{{{\bf{y}}^H}{{\bf{X}}_{\bf b}}{{\left( {{{\bf{X}}_{\bf b}^H}{{\bf{X}}_{\bf b}} + {\sigma ^2}{\bf{C}}_{\bf{g}}^{ - 1}} \right)}^{ - 1}}{{\bf{X}}_{\bf b}^H}{\bf{y}}}$. To handle this, we propose to establish an appropriate upper bound to the function $\Phi({\bf b})$ and then solve the corresponding surrogate problem in each ADMM iteration.
							
							\begin{proposition}
								Letting ${{\bf w}_{\bf b}} \triangleq {{\bf{X}}_{\bf b}^H}{\bf{y}} $ and ${{\bf R}_{\bf b}} \triangleq {{{\bf{X}}_{\bf b}^H}{{\bf{X}}_{\bf b}} + {\sigma ^2}{\bf{C}}_{\bf{g}}^{ - 1}}$, a surrogate problem of subproblem (\ref{eq20a}) is given by
								\begin{equation}
									\begin{small}
										\begin{aligned}	
											{{\bf{b}}^{n}} = \mathop {\arg \min }\limits_{{\bf{b}} \in {{[0,1]}^N}} {\lambda ^{n - 1}}\left\| f({\bf{b}}) \right\|_F^2 - 2{\mathop{\rm Re}\nolimits} \left\{ {{\rm{tr}}\left( {{f({\bf{b}})}^H} {{\bf D}^{n-1}}\right)} \right\} \\
											- \alpha \left\| {{\bf{b}} - {{\bf{0.5}}_N}} \right\|_2^2 + \frac{\mu }{2}{\left\| {\bf A}{\bf b}+ {{\bf{z}}^{n-1}} - {\bm \theta} + {{\bm \eta}^{n-1}} \right\|_2^2}, \label{eq25}
										\end{aligned}
									\end{small}
								\end{equation}
								where ${{\bf D}^{n-1}} = {\left( {\lambda ^{n - 1}}{{\bf{I}}_{{N_t}}} - {{\bf{V}}_{{{\bf{b}}^{n-1}}}^H}{{\bf{V}}_{{{\bf{b}}^{n-1}}}}  \right)f({{\bf b}^{n-1}})} + {{\bf{V}}_{{{\bf{b}}^{n-1}}}^H}{{\bf Y}_D}$ and ${{\bf{V}}_{{{\bf{b}}^{n-1}}}} = {\rm {unvec}}\left({\bf{R}}_{{{\bf{b}}^{n-1}}}^{ - 1}{{\bf{w}}_{{{\bf{b}}^{n-1}}}}\right)$ with ${\rm{unvec}}(\cdot)$ denoting the inverse vectorization operation.
							\end{proposition}
							\begin{IEEEproof}
								See Appendix \ref{Ap_2}.	
							\end{IEEEproof}
							
							Recall that $\bf A$ is column-wise orthogonal, which indicates that ${{\bf A}^T}{\bf A}$ is a diagonal matrix. Denote the $i$-th diagonal element of ${{\bf A}^T}{\bf A}$ by $\Lambda_i$. Thus, problem (\ref{eq25}) can be equivalently decomposed into ${N_t}{T_D}$ parallel subproblems, which take the forms of (\ref{eq26}) at the top of the page, where $d_{kt}^{n - 1}$ is the $(k,t)$-th entry of ${\bf D}^{n-1}$, ${b_{i^{t,k,q}}}$ is the $q$-th element of ${{{{\tilde{\bf b}}}^{t,k}}}$ with $i^{t,k,q} \triangleq Q[{N_t}(t - 1) + k - 1] + q$, and ${\bf{a}}_{{i^{t,k,q}}}$ is the $i^{t,k,q}$-th column of $\bf A$. 
							%Then, by taking the real and imaginary part of $f({{{{\bf{\tilde b}}}^{t,k}}})$, respectively, each subproblem in (\ref{eq26}) can be further decomposed into
							%\begin{equation}
							%	\begin{aligned}
								%	{\left({{{\bf{\tilde b}}}^{t,k}}\right)^n} = \mathop {\arg \min }\limits_{{{{\bf{\tilde b}}}^{t,k}} \in {{[0,1]}^Q}} &{\lambda ^{n - 1}}{\left| {f({{{\bf{\tilde b}}}^{t,k}})} \right|^2} - 2{\rm{Re}}\left\{ {{{\left( {f({{{\bf{\tilde b}}}^{t,k}})} \right)}^*}s_{kt}^{n - 1}} \right\} - \alpha \sum\limits_{l = 1}^Q {\left( {{b_{{i^{t,k,q}}}^2} - {b_{{i^{t,k,q}}}}} \right)}  \\ &+ \frac{\mu }{2}\sum\limits_{l = 1}^Q {\left( {d_{i^{t,k,q}}}{b_{{i^{t,k,q}}}^2 + 2{\bf{a}}_{{i^{t,k,q}}}^T\left( {{{\bf{z}}^{n - 1}} - {\bf{\theta }} + {{\bf{\eta }}^{n - 1}}} \right){b_{{i^{t,k,q}}}}} \right)},\label{eq27}
								%	\end{aligned}
							%\end{equation}
							We notice that each of the resulting subproblems only involves $Q$ variables and can therefore be readily tackled based on the block coordinate descent (BCD) method. Concretely, by regarding other variables as constants, we take the derivative of the objective function in (\ref{eq26}) w.r.t. each ${b_{i^{t,k,q}}}$ and set it to zero, which yields
							\setcounter{equation}{20} 
							\begin{equation}
								%			\begin{small}
									\begin{aligned}
										\frac{{\partial \phi ({{{\tilde{\bf b}}}^{t,k}})}}{{\partial {b_{{i^{t,k,q}}}}}} - \alpha &\left( {2{b_{{i^{t,k,q}}}} - 1} \right) + \mu \big( {\Lambda_{{i^{t,k,q}}}}{b_{{i^{t,k,q}}}} \\ & + {\bf{a}}_{{i^{t,k,q}}}^T\left( {{{\bf{z}}^{n - 1}} - {\bm{\theta }} + {{\bm{\eta }}^{n - 1}}} \right) \big) = 0,\label{eq27}
									\end{aligned}
									%		\end{small}
							\end{equation}
							where ${\phi({{{\tilde{\bf b}}}^{t,k}})}={\lambda ^{n - 1}}{\left| {f({{{\tilde{\bf b}}}^{t,k}})} \right|^2} - 2{\rm{Re}}\left\{ {{{\left( {f({{{\tilde{\bf b}}}^{t,k}})} \right)}^*}d_{kt}^{n - 1}} \right\}$. For simplicity, we temporarily denote $\frac{{\partial \phi ({{{\bf{\tilde b}}}^{t,k}})}}{{\partial {b_{{i^{t,k,q}}}}}}$ in the form of ${{\beta}^{n-1}_{{i^{t,k,q}}}}{b_{{i^{t,k,q}}}}+{{\gamma}^{n-1}_{{i^{t,k,q}}}}$, where the detailed expressions of ${{\beta}^{n-1}_{{i^{t,k,q}}}}$ and ${{\gamma}^{n-1}_{{i^{t,k,q}}}}$ for QPSK and 16QAM are provided in Appendix \ref{Ap_3}. Consequently, a closed-form solution to each ${b_{i^{t,k,q}}}$ can be obtained as
							\begin{equation}
								\begin{small}
									\begin{aligned}
										b_{{i^{t,k,q}}}^n = \prod\nolimits_{\left[0,1\right]} {\left( {\frac{{\mu {\bf{a}}_{{i^{t,k,q}}}^T\left( {{\bm{\theta }} - {{\bf{z}}^{n - 1}} - {{\bm{\eta }}^{n - 1}}} \right) - {\gamma^{n-1}_{{i^{t,k,q}}}} - \alpha }}{{\mu {\Lambda_{{i^{t,k,q}}}} + {\beta^{n-1}_{{i^{t,k,q}}}} - 2\alpha }}} \right)},\label{eq28}
									\end{aligned}
								\end{small}
							\end{equation}
							where ${\prod\nolimits_{{\left[0,1\right]}}}(\cdot)$ means projecting the input onto the interval $\left[0,1\right]$. We note that for QPSK, the $Q = 2$ bit variables in subproblem (\ref{eq26}) can be perfectly decoupled by invoking the bit-to-symbol mapping function in (\ref{eq9}), which means that (\ref{eq28}) is the optimal solution to subproblem (\ref{eq26}).
							
							\begin{algorithm}[t]
								\small
								\caption{JCDD Receiving Algorithm for Gaussian MIMO Channels}
								\begin{algorithmic}[1]
									\REQUIRE $\bf P$, ${\bf Y}_P$, ${\bf Y}_D$, $\bf{A}$, $\bm{\theta}$, $\bf{H}$, and $I_\text{max}$.%The received signals $ {{\bf{y}}}$, the imperfect CSI ${\hat{\bf h}}$, the  covariance matrix of CSI uncertainty ${{\bf{\Sigma }}_{\bf{h}}}$ and the maximum number of iteration $I_{max}$.
									\STATE Initialize: ${\bf{b}}^{0}$, ${\bf{z}}^{0}$, ${\bm{\eta}}^{0}$, %$c^0$, 
									$n = 0$.
									\REPEAT 
									\STATE $n \gets n+1$.
									\FOR {$k=1,2,\cdots,{N_t},t=1,2,\cdots,{T_D}$}
									\STATE Sequentially update $b_{{i^{t,k,q}}}^n$ for $q=1,2,\cdots,Q$ via (\ref{eq28}).
									\ENDFOR
									\STATE Update ${\bf{z}}^{n}$ via (\ref{eq20b}).
									\STATE Update ${\bm{\eta }}^{n}$ via (\ref{eq20c}).
									%		\STATE ${c^{n}}\gets \frac{{1 + \sqrt {1 + 4{\left(c^{n - 1}\right)^2}} }}{2}$.
									%		\STATE ${\bf{z}}^{n} \gets {\bf{z}}^{n} + \frac{c^{n - 1}-1}{c^{n}}\left({\bf{z}}^{n} - {\bf{z}}^{n-1}\right)$.
									%		\STATE ${\bm{\eta}}^{n} \gets {\bm{\eta}}^{n} + \frac{c^{n - 1}-1}{c^{n}}\left({\bm{\eta}}^{n} - {\bm{\eta}}^{n-1}\right)$.
									\STATE ${\hat{b}}_i \gets 0$ if $b_i^{n}<0.5$, or ${\hat{ b}}_i \gets 1$ otherwise, for $i=1,2,\cdots,N$.
									\UNTIL{${\left[\sum\limits_{i = 1}^N {{H_{ji}}}{\hat{b}_i}\right]_2} = 0$ for $j=1,2,\cdots,M$ \textbf{or} $n = I_{\text{max}}$.}
									%\STATE Calculate $\hat{\bf g}={\left( {{{\bf X}_{\hat{\bf b}}^H}{\bf X}_{\hat{\bf b}} + {\sigma ^2}{\bf{C}}_{\bf{g}}^{ - 1}} \right)^{ - 1}}{{\bf X}_{\hat{\bf b}}^H}{\bf{y}}$, where ${\bf X}_{\hat{\bf b}} ={\left[ {{\bf{P}},f({\hat{\bf b}})} \right]^T}\otimes{{\bf I}_{N_r}}$.
									\ENSURE  ${\hat{\bf b}}$ and $\hat{\bf g}={\left( {{{\bf X}_{\hat{\bf b}}^H}{\bf X}_{\hat{\bf b}} + {\sigma ^2}{\bf{C}}_{\bf{g}}^{ - 1}} \right)^{ - 1}}{{\bf X}_{\hat{\bf b}}^H}{\bf{y}}$, where ${\bf X}_{\hat{\bf b}} ={\left[ {{\bf{P}},f({\hat{\bf b}})} \right]^T}\otimes{{\bf I}_{N_r}}$.
								\end{algorithmic}
								\label{JCDD-G}
							\end{algorithm}	
							
							Overall, the proposed ADMM-based JCDD receiving algorithm is summarized in Algorithm \ref{JCDD-G}. Specifically, 
							%line 9-11 correspond to the predictor-corrector-type acceleration steps for fast convergence of ADMM \cite{Fast5456443}, 
							line 10 means early terminating the iterations once the parity check constraints are all satisfied. %,  and line 11 follows from (\ref{eq16}) by invoking the final codeword estimate $\hat{\bf b}$. 
							Note that to avoid the high computational cost required in a complete BCD process, we only perform (\ref{eq28}) once for each ${b_{i^{t,k,q}}}$ in each ADMM iteration. Although this results in an inexact solution to problem (\ref{eq20a}) and thus renders the convergence analysis of ADMM quite challenging, the empirical results suggest that Algorithm 1 can still converge with a satisfactory performance. 
							
							\subsection{Network Design}
							For Algorithm \ref{JCDD-G}, the matrix inversion ${\bf{R}}_{{{\bf{b}}^{n-1}}}^{ - 1}$ involved in the update of ${\bf b}^{n}$ via (\ref{eq28}) requires a high complexity up to ${\cal O}\left({N_r^3}{N_t^3}\right)$ in each ADMM iteration, which is unfavorable to the design of an efficient neural network. Fortunately, it is worthwhile noting that when the channel coefficients are i.i.d. Gaussian distributed with zero mean and variance $\sigma^2_{g}$, i.e., ${\bf {C_g}} = {\sigma^2_{g}}{{\bf I}}_{{N_r}{N_t}}$, we have
							\begin{equation}
								%	\begin{small}
									\begin{aligned}
										{\bf{R}}_{{{\bf{b}}^{n-1}}}^{ - 1}
										%=&{\left({{\bf{X}}_{{\bf b}^{n-1}}^H}{{\bf{X}}_{{\bf b}^{n-1}}} + {\sigma ^2}{\bf {C}}_{\bf g}^{-1}\right)^{ - 1}}\\
										%\mathop = \limits^{(a)}&\left(\left({{{\bf{S}}_{{\bf b}^{n-1}}^{*}}{{\bf{S}}_{{\bf b}^{n-1}}^T}}\right)\otimes{{\bf I}_{N_r}}+\frac{\sigma ^2}{\sigma^2_{g}}{{\bf I}}_{{N_r}{N_t}}\right)^{-1}\\
										\mathop = \limits^{(a)}\left({{{\bf{S}}_{{\bf b}^{n-1}}^{*}}{{\bf{S}}_{{\bf b}^{n-1}}^T}} + \frac{\sigma ^2}{\sigma^2_{g}}{{\bf I}}_{N_t}\right)^{-1} \otimes {{\bf I}}_{N_r},
										\label{eq4C1}
									\end{aligned}
									%	\end{small}
							\end{equation}
							where ${\bf{S}}_{{\bf b}^{n-1}} = \left[ {{\bf{S}}_P,f({\bf b}^{n-1})} \right]$, (a) is achieved by utilizing $\left({\bf A} \otimes {\bf B}\right)\left({\bf C} \otimes {\bf D}\right) = \left({\bf A}{\bf C}\right) \otimes \left({\bf B}{\bf D}\right)$, $\left({\bf A} \otimes {\bf B}\right) + \left({\bf C} \otimes {\bf B}\right) = \left({\bf A}+{\bf C}\right) \otimes {\bf B}$, and $\left({\bf A} \otimes {\bf B}\right)^{-1} = {\bf A}^{-1} \otimes {\bf B}^{-1}$. In this way, the complexity of calculating ${\bf{R}}_{{{\bf{b}}^{n-1}}}^{ - 1}$ can be reduced to ${\cal O}\left({N_t^3}\right)$.  
							Moreover, it follows from (\ref{eq4C1}) that   
							\begin{equation}
								%				\begin{small}
									\begin{aligned}
										&{{\bf{v}}_{{{\bf{b}}^{n-1}}}}={\bf{R}}_{{{\bf{b}}^{n-1}}}^{ - 1}{{\bf{w}}_{{{\bf{b}}^{n-1}}}}\\
										=& \left(\left({{{\bf{S}}_{{\bf b}^{n-1}}^{*}}{{\bf{S}}_{{\bf b}^{n-1}}^T}} + \frac{\sigma ^2}{\sigma^2_{g}}{{\bf I}}_{N_t}\right)^{-1} \otimes {{\bf I}}_{N_r}\right) \left({{\bf{S}}_{{\bf b}^{n-1}}^{*}}\otimes {{\bf I}}_{N_r} \right) {\bf{y}}\\
										\mathop = \limits^{(a)}& \left(\left(\left({{{\bf{S}}_{{\bf b}^{n-1}}^{*}}{{\bf{S}}_{{\bf b}^{n-1}}^T}} + \frac{\sigma ^2}{\sigma^2_{g}}{{\bf I}}_{N_t}\right)^{-1} {{\bf{S}}_{{\bf b}^{n-1}}^{*}} \right)  \otimes {{\bf I}}_{N_r}\right) {\bf{y}},		
										\label{eq4C2}
									\end{aligned}
									%			\end{small}
							\end{equation}
							where (a) is also obtained based on $\left({\bf A} \otimes {\bf B}\right)\left({\bf C} \otimes {\bf D}\right) = \left({\bf A}{\bf C}\right) \otimes \left({\bf B}{\bf D}\right)$. Then, by inversely applying ${\rm{vec}}\left({\bf A}{\bf B}{\bf C}\right)=\left({{\bf C}^T} \otimes {\bf A}\right){\rm{vec}}\left({\bf B}\right)$, it further yields a more concise expression of ${{\bf{V}}_{{{\bf{b}}^{n-1}}}} = {\rm {unvec}}\left({{\bf{v}}_{{{\bf{b}}^{n-1}}}}\right)$ as
							\begin{equation}
								%		\begin{small}
									%	\begin{aligned}
										{{\bf{V}}_{{{\bf{b}}^{n-1}}}} = {\bf Y} {{\bf{S}}_{{\bf b}^{n-1}}^{H}} \left({{{\bf{S}}_{{\bf b}^{n-1}}}{{\bf{S}}_{{\bf b}^{n-1}}^H}} + \frac{\sigma ^2}{\sigma^2_{g}}{{\bf I}}_{N_t}\right)^{-1},		
										\label{eq4C3}
										%	\end{aligned}
									%\end{small}
								\end{equation}
								which therefore simplifies the calculation of ${\bf D}^{n-1}$ and, furthermore, the update of ${\bf b}^{n}$. 
								
								Inspired by the complexity advantage of the above-mentioned i.i.d. case, we consider to construct an upper bound to $\Phi({\bf b})$ in an i.i.d.-like form by		
								\begin{equation}
									\begin{aligned}
										\Phi({\bf b})&=-{{{\bf{y}}^H}{{\bf{X}}_{\bf b}}{{\left( {{{\bf{X}}_{\bf b}^H}{{\bf{X}}_{\bf b}} + {\sigma ^2}{\bf{C}}_{\bf{g}}^{ - 1}} \right)}^{ - 1}}{{\bf{X}}_{\bf b}^H}{\bf{y}}} \\
										&\leq  -{{{\bf{y}}^H}{{\bf{X}}_{\bf b}}{{\left( {{{\bf{X}}_{\bf b}^H}{{\bf{X}}_{\bf b}} + \upsilon{\bf I}_{{N_r}{N_t}}} \right)}^{ - 1}}{{\bf{X}}_{\bf b}^H}{\bf{y}}},
										\label{eq4C4}
									\end{aligned}
								\end{equation}
								which holds when $\upsilon$ satisfies $\upsilon{\bf I}_{{N_r}{N_t}} \succeq {\sigma ^2}{\bf{C}}_{\bf{g}}^{-1}$. Note that we can set ${\upsilon}={\lambda}_{\rm {max}}\left({\sigma ^2}{\bf{C}}_{\bf{g}}^{-1}\right)$. Thus, after replacing ${\bf R}_{\bf b}$ in (\ref{eq21}) with ${\dot{\bf R}}_{\bf b} \triangleq {{{\bf{X}}_{\bf b}^H}{{\bf{X}}_{\bf b}} + \upsilon{{\bf I}}_{{N_r}{N_t}}}$, we can directly follow the previous derivation and finally arrive at a similar result to (\ref{eq28}). 
								
								%Now we are ready to develop a model-driven neural network based on the deep unfolding technique. 
								It is known that the performance and the convergence rate of the ADMM algorithm highly depend on the penalty parameters. To avoid an exhaustive search, we unfold the ADMM iterations of Algorithm 1 into ${L_\text{max}}$ network layers and set the involved penalty parameters to be trainable so that they can be optimized via offline training. Specifically, we use different penalty parameters for each layer, denoted by $\{{\mu^l},{\alpha^l}\}_{l = 1}^{L_\text{max}}$, to provide more degrees of freedom and enhance the expressiveness of the network. In addition, note that although $\lambda^{n-1}$ involved in the calculation of ${{\beta}^{n-1}_{{i^{t,k,q}}}}$ and ${{\gamma}^{n-1}_{{i^{t,k,q}}}}$ in (\ref{eq28}) and $\upsilon$ introduced in (\ref{eq4C4}) can be determined using some traditional ways, e.g., ${\lambda^{n-1}} = {\lambda}_{\rm {max}}\left({{\bf{V}}_{{{\bf{b}}^{n-1}}}^H}{{\bf{V}}_{{{\bf{b}}^{n-1}}}}\right)$ and ${\upsilon}={\lambda}_{\rm {max}}\left({\sigma ^2}{\bf{C}}_{\bf{g}}^{-1}\right)$, the operation of extracting the maximum eigenvalue is nondifferentiable, which impedes the training process. Therefore, they are also recast into layer-wise trainable parameters. In particular, instead of directly sharing the same parameters for different channel realizations, we incorporate the channel information into ${\lambda^l}$ and  ${\upsilon^l}$ by rewriting them as $ {o_{\lambda}^l}{\lambda}_{\rm {max}}\left({{\bf{V}}_{{{\bf{b}}^{0}}}^H}{{\bf{V}}_{{{\bf{b}}^{0}}}}\right)$ and $ {o_{\upsilon}^l}{\lambda}_{\rm {max}}\left({\sigma ^2}{\bf{C}}_{\bf{g}}^{-1}\right)$, respectively. In this way, only the corresponding adjustment factors $\{{o_{\lambda}^l}\}_{l = 1}^{L_\text{max}}$ and $\{{o_{\upsilon}^l}\}_{l = 1}^{L_\text{max}}$ need to be trained, which can lower the training difficulty as well as improving the adaptivity of the network.
								%\footnote{To facilitate the software tools such as Tensorflow, the complex variables should be transformed into the equivalent real domain, which is a common process and thus omitted here due to the space limitation.}
								Accordingly, the $l$-th layer of the proposed network can be constructed by
								\begin{subequations}
									\label{eq4C5}
									\begin{equation}
										%				\begin{small}
											\begin{aligned}
												{{\bf{V}}_{{{\bf{b}}^{l-1}}}} = {\bf Y}{{\bf{S}}_{{\bf b}^{n-1}}^{H}} \left({{\bf{S}}_{{\bf b}^{n-1}}{{\bf{S}}_{{\bf b}^{n-1}}^H}} + {o_{\upsilon}^l}{\lambda}_{\rm {max}}\left({\sigma ^2}{\bf{C}}_{\bf{g}}^{-1}\right){{\bf I}}_{N_t}\right)^{-1}, \label{eq4C5_a}
											\end{aligned}
											%\end{small}
										\end{equation}
										\begin{equation}
											%				\begin{small}
												\begin{aligned}
													{{\bf S}^{l-1}} = &\big( {o_{\lambda}^{l}}{\lambda}_{\rm {max}}\left({{\bf{V}}_{{{\bf{b}}^{0}}}^H}{{\bf{V}}_{{{\bf{b}}^{0}}}}\right){{\bf{I}}_{{N_t}}} - {{\bf{V}}_{{{\bf{b}}^{l-1}}}^H}{{\bf{V}}_{{{\bf{b}}^{l-1}}}}  \big)f({{\bf b}^{l-1}}) \\&+ {{\bf{V}}_{{{\bf{b}}^{l-1}}}^H}{{\bf Y}_D}, \label{eq4C5_b}
												\end{aligned}
												%\end{small}
											\end{equation}
											\begin{equation}
												%			\begin{small}
													\begin{aligned}
														{{\bf{b}}}^{l}\left[{\cal I}_q\right] =\prod\nolimits_{\left[0,1\right]^{N/Q}} \Big(\big({\mu^l} {\bf{A}}\left[:,{\cal I}_q\right]^T\left({{\bm{\theta }} - {{\bf{z}}^{l - 1}} - {{\bm{\eta }}^{l - 1}}} \right) \\- {{\bm \gamma}^{l-1}}\left[{\cal I}_q\right] - {\alpha^l}{{\bf{1}}_{N/Q}} \big) \oslash {\bm \delta}^l\left[{\cal I}_q\right]\Big), \forall q, \label{eq4C5_c}
													\end{aligned}
													%\end{small}
												\end{equation}
												%	\begin{equation}
													%	{\bf{b}}^{l}\left[{\cal I}_q\right] = {\rm{ReLU}}\left({\bar{\bf{b}}}^{l}\left[{\cal I}_q\right] \right)-{\rm{ReLU}}\left({\bar{\bf{b}}}^{l}\left[{\cal I}_q\right]-{{\bf{1}}_{N/Q}} \right),q=1,2,\cdots,Q, \label{eq4C5_d}
													%\end{equation}
													\begin{equation}
														%				\begin{small}
															\begin{aligned}
																{\bf{b}}^{l} =  {\rm{Concat}}\left({\bf{b}}^{l}\left[{\cal I}_q\right],\forall q\right), \label{eq4C5_e}
															\end{aligned}
															%\end{small}
														\end{equation}
														\begin{equation}
															%				\begin{small}
																\begin{aligned}
																	{ {\bf{z}}}^{l} = {\rm{ReLU}}\left({\bm \theta} - {\bf A}{{\bf b}^{l}}- {\bm{\eta}}^{l-1}\right),\label{eq4C5_f}
																\end{aligned}
																%\end{small}
															\end{equation}
															\begin{equation}
																%			\begin{small}
																	\begin{aligned}
																		{ {\bm{\eta }}}^{l} = {\bm{\eta }}^{l-1} + {\bf A}{{\bf b}^{l}}+ {  {\bf{z}}}^{l} - {\bm \theta}, \label{eq4C5_g}
																	\end{aligned}
																	%\end{small}
																\end{equation}
								\end{subequations}
								
								where ${\bm \delta}^l = {\mu^l}{{\rm diag}\left({{\bf A}^T}{\bf A}\right)} + {{\bm \beta}^{l-1}} - 2{\alpha^l}{{\bf{1}}_{N}}$ with ${\rm {diag}}(\cdot)$ being the diagonal vector of the input matrix, $\oslash$ denotes the element-wise division, and ${\cal I}_q$ is defined as the index set of the $q$-th bit variables corresponding to all symbols, i.e, ${\cal I}_q \triangleq \{ {i^{t,k,q}}\left| {\forall t,\forall k} \right.\}$ with $i^{t,k,q} \triangleq Q[{N_t}(t - 1) + k - 1] + q$. Note that ${\bm \beta}^{l-1}$ and ${\bm \gamma}^{l-1}$ involved in (\ref{eq4C5_c}) can be obtained using ${\lambda}^{l-1} = {o_{\lambda}^{l-1}}{\lambda}_{\rm {max}}\left({{\bf{V}}_{{{\bf{b}}^{0}}}^H}{{\bf{V}}_{{{\bf{b}}^{0}}}}\right)$, ${{\bf S}^{l-1}}$, and $f({{\bf b}^{l-1}})$ according to Appendix \ref{Ap_3}. Besides, ${\rm{ReLU}}(\cdot)$ and ${\rm{Concat}}(\cdot)$ represent the rectified linear unit (ReLU) activation function and the concatenating function, respectively.  ${\prod\nolimits_{{\left[0,1\right]^{N/Q}}}}(\cdot)$ means projecting each entry of the input vector onto the interval $\left[0,1\right]$.
															
								Note that some variants of ADMM can enjoy a faster convergence rate, such as the relaxed and accelerated ADMM (R-A-ADMM) in \cite{Fast5456443}, which combines the over-relaxation scheme with a predictor-corrector-type acceleration step. Inspired by this, we improve the network structure and additionally introduce trainable parameters for each layer, which takes the form:
													\begin{subequations}
	\label{eq4C7}
	\begin{equation}
		%				\begin{small}
			\begin{aligned}
				\text{(\ref{eq4C5_a})},\quad \text{(\ref{eq4C5_b})},\quad \text{(\ref{eq4C5_c})},\quad \text{(\ref{eq4C5_e})},\label{eq4C7_a}
			\end{aligned}
			%\end{small}
		\end{equation}
		%	\begin{equation}
			%		{{\bf{b}}}^{l}\left[{\cal I}_q\right] =\prod\nolimits_{\left[0,1\right]^{N/Q}} \left( \left({{\mu^l} {\bf{A}}\left[:,{\cal I}_q\right]^T\left( {{\bm{\theta }} - {\hat{\bf{z}}^{l - 1}} - {\hat{\bm{\eta }}^{l - 1}}} \right) - {{\bm \gamma}^{l-1}}\left[{\cal I}_q\right] - {\alpha^l}{{\bf{1}}_{N/Q}} }\right) \oslash {\bm \delta}^l\left[{\cal I}_q\right]\right), \forall q, \label{eq4C7_b}
			%	\end{equation}
		%	\begin{equation}
			%	{\bf{b}}^{l}\left[{\cal I}_q\right] = {\rm{ReLU}}\left({\bar{\bf{b}}}^{l}\left[{\cal I}_q\right] \right)-{\rm{ReLU}}\left({\bar{\bf{b}}}^{l}\left[{\cal I}_q\right]-{{\bf{1}}_{N/Q}} \right),q=1,2,\cdots,Q, \label{eq4C5_d}
			%\end{equation}
			%	\begin{equation}
				%		\text{(\ref{eq4C5_e})}, \label{eq4C7_e}
				%	\end{equation}
			\begin{equation}
				%				\begin{small}
					\begin{aligned}
						{\bar{\bf{z}}}^{l} = {\rm{ReLU}}\left({\bm \theta} - {o_r^l}{\bf A}{{\bf b}^{l}}-\left(1-{o_r^l}\right)\left({\bm \theta}-{\bf{z}}^{l-1}\right) - {\bm{\eta}}^{l-1}\right),\label{eq4C7_b}
					\end{aligned}
					%\end{small}
				\end{equation}
				\begin{equation}
					%				\begin{small}
						\begin{aligned}
							{\bar {\bm{\eta }}}^{l} = {\bm{\eta }}^{l-1} + {o_r^l}{\bf A}{{\bf b}^{l}}+\left(1-{o_r^l}\right)\left({\bm \theta}-{\bf{z}}^{l-1}\right)+ {\bar  {\bf{z}}}^{l} - {\bm \theta}, \label{eq4C7_c}
						\end{aligned}
						%\end{small}
					\end{equation}
					\begin{equation}
						%					\begin{small}
							\begin{aligned}
								{\bf{z}}^{l} = {\bar{\bf{z}}}^{l} + {o_p^l}\left({\bar {\bf{z}}}^{l} - {\bar{\bf{z}}}^{l-1}\right), \label{eq4C7_d}
							\end{aligned}
							%	\end{small}
					\end{equation}
					\begin{equation}
						%					\begin{small}
							\begin{aligned}
								{\bm{\eta}}^{l} =  {\bar{\bm{\eta }}}^{l} + {o_p^l}\left({\bar{\bm{\eta }}}^{l} - {\bar{\bm{\eta}}}^{l-1}\right), \label{eq4C7_e}
							\end{aligned}
							%	\end{small}
					\end{equation}
				\end{subequations} 
				where $o_r^l$ and $o_p^l$ denote the trainable relaxation parameter and predictor-corrector parameter, respectively. %Note that (\ref{eq4C7}) can be degenerated to (\ref{eq4C5}) when $o_r^l = 1$ and $o_p^l = 0$, indicating the introduction of them at least incurs no performance loss.
				Furthermore, to simplify the expressions, we let ${\bf{w}}^{l} = {\bm \theta} - {o_r^l}{\bf A}{{\bf b}^{l}}-\left(1-{o_r^l}\right)\left({\bm \theta}-{\bf{z}}^{l-1}\right) - {\bm{\eta}}^{l-1}$ and recast (\ref{eq4C7_b})-(\ref{eq4C7_e}) by
				\begin{subequations}
					\label{eq4C8}
					\begin{equation}
						%						\begin{small}
							\begin{aligned}
								{\bf{z}}^{l} = {\rm{ReLU}}\left({\bf{w}}^{l}\right) + {o_p^l}\left({\rm{ReLU}}\left({\bf{w}}^{l}\right) - {\rm{ReLU}}\left({\bf{w}}^{l-1}\right)\right), \label{eq4C8_d} 
							\end{aligned}
							%	\end{small}
					\end{equation}
					\begin{equation}
						%						\begin{small}
							\begin{aligned}
								{\bm{\eta}}^{l} =  {\bf{z}}^{l} - \left(1+{o_p^l}\right){\bf{w}}^{l} + {o_p^l}{\bf{w}}^{l-1}. \label{eq4C8_e}
							\end{aligned}
							%	\end{small}
					\end{equation}
				\end{subequations} 
				
				\begin{figure}[!t]
					\centering
					\includegraphics[width=8cm]{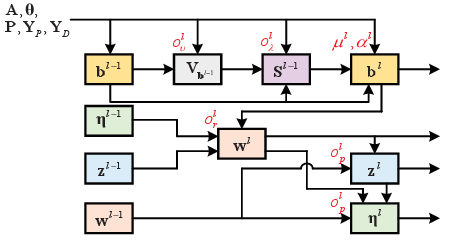}
					\caption{Block diagram of the $l$-th layer  of JCDDNet-G.}
					\label{fig2}
				\end{figure}
				The block diagram of the $l$-th layer of the proposed JCDDNet-G is illustrated in Fig. \ref{fig2}.
				
				\section{JCDD for Sparse mmWave MIMO Channels}
				The mmWave MIMO systems can promise low transmission latency owing to the massive bandwidth, which are thereby attractive for URLLC. However, it has been verified that the energy of an mmWave MIMO channel is mainly dominated by a limited number of scattering clusters, resulting in its non-Gaussian distribution and the sparsity in the beamspace domain. Therefore, the previously proposed JCDDNet-G is no longer suitable. To fill the gap, we establish another JCDD receiver for sparse mmWave MIMO channels in this section. Similarly to Section III, we also start by formulating a MAP-JCDD problem and then solve it based on the ADMM framework. Furthermore, the corresponding deep unfolding network JCDDNet-S is provided.
				
				\subsection{Problem Formulation}
				{ We consider the uniform planar array (UPA) of size ${N^y}$ and ${N^z}$ and use the subscripts $r$ and $t$ to distinguish the receive and transmit antennas. To exploit the sparsity of mmWave MIMO channels, we first rewrite the channel matrix $\bf G$ using the beamspace-domain representation by
					\begin{equation}
						{\bf{G}} = {{\bf F}_r}{{\bf G}_s}{{\bf F}^H_t},\label{eq31}
				\end{equation}}
				where ${{\bf F}_r} = {{\bf F}_{N^y_r}} \otimes {{\bf F}_{N^z_r}}$ and ${{\bf F}_t}  = {{\bf F}_{N^y_t}} \otimes {{\bf F}_{N^z_t}}$, ${{\bf F}_{N^y_r}} \in {\mathbb{C}} ^{{N^y_r} \times {N^y_r}}$, ${{\bf F}_{N^z_r}} \in {\mathbb{C}} ^{{N^z_r} \times {N^z_r}}$, ${{\bf F}_{N^y_t}} \in {\mathbb{C}} ^{{N^y_t} \times {N^y_t}}$, and ${{\bf F}_{N^z_t}} \in {\mathbb{C}} ^{{N^z_t} \times {N^z_t}}$ are unitary discrete fourier transform (DFT) matrices. ${{\bf G}_s} \in {\mathbb{C}} ^{{N_r} \times {N_r}}$ denotes the sparse beamspace-domain channel matrix. Since the number of channel clusters $N_{cl}$ is relatively small in the mmWave band, i.e., ${N_{cl}} \ll {{N_r}{N_t}}$, there are only a few groups of large-amplitude entries in ${{\bf G}_s}$, each group corresponding to one channel cluster. We note that ${{\bf G}_s}$ is not perfectly ``sparse'' due to the power leakage, which means that other entries of ${{\bf G}_s}$ are of small amplitude but not exactly zero \cite{mmwave}. Therefore, we can assume that ${{\bf G}_s}$ follows a Laplace distribution, i.e.,
				\begin{equation}
					p({{\bf{G}}_s}) = {\left( {{{{\varepsilon ^2}} \over {2\pi }}} \right)^{{N_r}{N_t}}}\exp \left( { - \varepsilon {{\left\| {{{\bf{G}}_s}} \right\|}_1}} \right),\label{eq32}
				\end{equation}
				where $\varepsilon > 0$ is the sparsity parameter and  ${\left\| \cdot \right\|_1}$ denotes the sum of the absolute value of each entry of the input matrix. %the $\mathcal{L}_1$ norm of the input matrix. 
				
				%Since ${\bf F}_r$ is a unitary matrix, we plug (\ref{eq31}) into (\ref{eq6}) and utilize the unitary invariance of the Frobenius norm, which yields
				Let us plug (\ref{eq31}) into (\ref{eq5}) and obtain
				\begin{equation}
					p\left( {{\bf Y}\left| {{\bf{G}},{\bf S}} \right.} \right) = \frac{1}{{{\left(\pi{\sigma^2}\right) ^{{N_r}T}}}}\exp \left( { - \frac{{\left\| {{{\bf{F}}_r^H}{\bf Y} - {{{\bf{G}}_s}{\bf{F}}_t^H}{\bf S}} \right\|_F^2}}{{{\sigma ^2}}}} \right),\label{eq33}
				\end{equation}  
				where ${\left\| \cdot \right\|_F}$ denotes the Frobenius norm of the input matrix and we utilize the fact that  ${\bf F}_r$ is a unitary matrix and the unitary invariance of the Frobenius norm. Then, by substituting (\ref{eq32}) and (\ref{eq33}) into (\ref{eq4}) with some algebraic simplifications and considering the bit-to-symbol mapping function as well as the LDPC code constraints, the MAP-JCDD problem for sparse mmWave MIMO channels can be formulated by
				\begin{equation} 
					\begin{aligned}
						\mathop {\min }\limits_{{{\bf{G}}_s},{\bf b}}\quad&\left\| {{\bf{F}}_r^H{\bf Y} - {{\bf{G}}_s}{\bf{F}}_t^H{\bf S}_{\bf b}} \right\|_F^2 + {\sigma ^2}\varepsilon {\left\| {{{\bf{G}}_s}} \right\|_1}\\
						\text{s.t.}\quad&(1),(2).\label{eq34}
					\end{aligned}
				\end{equation}
				
				\subsection{ADMM-Based Solution}
				Similarly to Section~III-B, we also derive an ADMM-based algorithm to address problem (\ref{eq34}). Following the reformulation of problem (\ref{eq18}), problem (\ref{eq34}) can be further relaxed to
				\begin{equation} 
					\begin{aligned}
						\mathop {\min }\limits_{{{\bf{G}}_s},{\bf{b}} \in {{[0,1]}^N}} \quad&\Psi({{\bf G}_s},{\bf b}) -\kappa \left\| {{\bf{b}} - {\bf{0.5}}_N} \right\|_2^2\\
						\text{s.t.}\quad\quad&{\bf A}{\bf b}+ {\bf u}= {\bm \theta},\quad{\bf u}\geq{\bf 0},\label{eq35}
					\end{aligned}
				\end{equation}
				where $\Psi({{\bf G}_s},{\bf b}) = \left\| {{\bf{F}}_r^H{\bf Y} - {{\bf{G}}_s}{\bf{F}}_t^H{\bf S}_{\bf b}} \right\|_F^2 + {\sigma ^2}\varepsilon {\left\| {{{\bf{G}}_s}} \right\|_1}$,  ${\bf u} \in {\mathbb R}^{\Gamma^c}$ denotes the introduced auxiliary variable, and $-\kappa \left\| {{\bf{b}} - {\bf{0.5}}_N} \right\|_2^2$ is the penalty term with $\kappa > 0$ being the penalty parameter. Then, regarding $\left[{{\bf G}_s},{\bf b}\right]$ as a block, the augmented Lagrangian function w.r.t. (\ref{eq35}) is given by
				\begin{equation} 
					\begin{aligned}
						L_{\rho}\left(\left[{{\bf G}_s},{\bf b}\right],{\bf{u}},{\bm{\omega}}\right) =& \Psi({{\bf G}_s},{\bf b}) - \kappa \left\| {{\bf{b}} - {{\bf{0.5}}_N}} \right\|_2^2 \\&+ \frac{\rho }{2}{\left\| {\bf A}{\bf b}+ {\bf u} - {\bm \theta} + {\bm \omega} \right\|_2^2}- \frac{\rho }{2} {\left\| {\bm \omega}\right\|_2^2}, \label{eq36}
					\end{aligned}
				\end{equation}
				where ${\bm{\omega}} \in {\mathbb{R}}^{\Gamma^c}$ and ${\rho}>0$ represent the scaled dual variables and the penalty parameter, respectively. Thus, the $n$-th ADMM iteration can be written as \cite{ADMMBook}
				\begin{subequations}
					\label{eq37}
					\begin{equation}
						\begin{aligned}
							{\left[{{\bf G}_s},{\bf b}\right]^n} = &\mathop {\arg \min }\limits_{{{\bf{G}}_s},{\bf{b}} \in {{[0,1]}^N}} \Psi({{\bf G}_s},{\bf b}) - \kappa \left\| {{\bf{b}} - {{\bf{0.5}}_N}} \right\|_2^2 \\&+ \frac{\rho }{2}{\left\| {\bf A}{\bf b}+ {{\bf{u}}^{n-1}} - {\bm \theta} + {{\bm \omega}^{n-1}} \right\|_2^2}, \label{eq37a}
						\end{aligned}
					\end{equation}
					\begin{equation}
						{\bf{u}}^{n} = {\prod\nolimits_{{\left[0,\infty\right)}^{\Gamma^c}}}\left({\bm \theta} - {\bf A}{{\bf b}^{n}}- {\bm{\omega}}^{n-1}\right),\label{eq37b}
					\end{equation}
					\begin{equation}
						{\bm{\omega }}^{n} = {\bm{\omega }}^{n-1} + {\bf A}{{\bf b}^{n}}+ {{\bf u}^{n}} - {\bm \theta}.\label{eq37c}
					\end{equation}
				\end{subequations} 
				Note that subproblem (\ref{eq37a}) is non-convex w.r.t. $\left[{{\bf G}_s},{\bf b}\right]$ with ${\bf G}_s$ and $\bf b$ coupled in the objective function. To handle this, we adopt the BCD method to solve this subproblem by alternatively optimizing ${\bf G}_s$ and $\bf b$. 
				\subsubsection{Optimization of ${\bf G}_s$}
				For the $n$-th ADMM iteration, by fixing ${\bf b}$ with the last update ${\bf b}^{n-1}$, subproblem (\ref{eq37a}) becomes
				\begin{equation} 
					\begin{aligned}
						{{\bf G}_s^n} = \mathop {\arg \min }\limits_{{{\bf{G}}_s}} \left\| {{\bf{F}}_r^H{\bf Y} - {{\bf{G}}_s}{\bf{F}}_t^H{\bf S}_{{\bf b}^{n-1}} } \right\|_F^2 + {\sigma ^2}\varepsilon {\left\| {{{\bf{G}}_s}} \right\|_1},\label{eq38}
					\end{aligned}
				\end{equation}
				where ${\bf S}_{{\bf b}^{n-1}} = \left[ {{\bf{P}},{f({{\bf b}^{n-1}})}} \right]$. (\ref{eq38}) is convex w.r.t. ${\bf G}_s$ but its optimal closed-form solution is hard to be obtained. Considering that the term ${\sigma ^2}\varepsilon {\left\| {{{\bf{G}}_s}} \right\|_1}$ is non-differentiable, we propose to employ the proximal gradient descent (PGD) algorithm to iteratively solve problem (\ref{eq38}). Specifically, for a problem in the form of ${\bf{\hat x}} = \mathop {\arg \min }\limits_{\bf{x}} r({\bf{x}}) + q({\bf{x}})$, where $p({\bf{x}})$ is smooth, differentiable and convex while $q({\bf{x}})$ is convex but not necessarily differentiable, the PGD algorithm iteratively updates $\bf x$ by \cite{proximal}
				\begin{equation}
					{{\bf{x}}^i} = {{\rm{prox}}_q}\left( {{{\bf{x}}^{i - 1}} - \tau \nabla r\left( {{{\left( {{{\bf{x}}^*}} \right)}^{i - 1}}} \right)}; \tau \right),i = 1, 2, \cdots \label{eq39}
				\end{equation}
				where ${{\rm{prox}}_q}(\cdot)$ and $\tau$ denote the proximal operator of the function $q({\bf{x}})$ and the stepsize, respectively. We emphasize that the conjugate gradient $\nabla r\left( {{{\left( {{{\bf{x}}^*}} \right)}^{i - 1}}}\right)$ in (\ref{eq39}) is considered since the variables are complex-valued. Accordingly, letting $r({\bf G}_s) \triangleq \left\| {{\bf{F}}_r^H\left[ {{{\bf{Y}}_P},{{\bf{Y}}_D}} \right] - {{\bf{G}}_s}{\bf{F}}_t^H\left[ {{\bf{P}},{f({{\bf b}^{n-1}})}} \right]} \right\|_F^2$ and $q({\bf G}_s) \triangleq {\sigma ^2}\varepsilon {\left\| {{{\bf{G}}_s}} \right\|_1}$, the solution to problem (\ref{eq38}) can be obtained as 
				\begin{equation}
					\begin{small}
						\begin{aligned}
							{{\bf G}_s^n} &= {{\rm{prox}}_q}\left( {{{\bf G}_s^{n - 1}} - \tau \nabla r\left( {{{\left( {{{\bf G}_s^*}} \right)}^{n - 1}}} \right)}; \tau \right) \triangleq {{\rm{prox}}_q}\left( {{\bar{\bf G}}_s}; \tau \right), \label{eq40}
						\end{aligned}
					\end{small}
				\end{equation}
				where 
				\begin{equation}
					%			\begin{small}
						\begin{aligned}
							\nabla r\left( {{{\left( {{{\bf G}_s^*}} \right)}^{n - 1}}} \right) =\left( {{\bf{G}}_s^{n - 1}{\bf{F}}_t^H{\bf S}_{{\bf b}^{n-1}} - {\bf{F}}_r^H{\bf Y}} \right){\left( {{\bf{F}}_t^H{\bf S}_{{\bf b}^{n-1}}}\right)^H}, \label{eq41}
						\end{aligned}
						%		\end{small}
				\end{equation}
				and  
				\begin{equation}
					{{\rm{prox}}_q}\left( {{\bar{\bf G}}_s}; \tau \right) = \mathop {\arg \min }\limits_{\bf X} {\sigma ^2}\varepsilon {\left\| {{{\bar{\bf G}}_s}} \right\|_1} + \frac{1}{2\tau}\left\|{\bf X} - {{\bar{\bf G}}_s}\right\|_F^2, \label{eq42}
				\end{equation}
				whose solution can be expressed using the shrinkage operator as $h_{\rm {sk}}\left({\bar{\bf G}}_s;{\sigma ^2}\varepsilon\tau\right)$, which can be further rewritten in an element-wise form as
				\begin{equation}
					h_{\rm {sk}}\left({\bar{g}}_s;{\sigma ^2}\varepsilon\tau\right) = \frac{{\bar g}_s}{\left|{\bar g}_s\right|}\mathop {\max}\left\{|{\bar g}_s|-{\sigma ^2}\varepsilon\tau,0\right\}, \label{eq43}
				\end{equation}
				where ${\bar g}_s$ represents each entry of ${\bar{\bf G}_s}$. Note that we also only perform a one-step PGD update instead of iterating until convergence in each ADMM iteration. 
				
				\subsubsection{Optimization of $\bf b$} Given ${{\bf G}_s^n}$ and some mathematical manipulations, we can obtain the following problem w.r.t. $\bf b$:
				\begin{equation}
					\begin{aligned}
						{{\bf{b}}^{n}} = \mathop {\arg \min }\limits_{{\bf{b}} \in {{[0,1]}^N}} \left\| {{{\bf{Y}}_D} - {{\bf{G}}^n}{f({{\bf b}})}} \right\|_F^2 - \kappa \left\| {{\bf{b}} - {{\bf{0.5}}_N}} \right\|_2^2 \\ + \frac{\rho }{2}{\left\| {\bf A}{\bf b}+ {{\bf{u}}^{n-1}} - {\bm \theta} + {{\bm \omega}^{n-1}} \right\|_2^2},\label{eq44}
					\end{aligned}
				\end{equation}
				where ${{\bf{G}}^n} = {{\bf{F}}_r}{{\bf{G}}_s^n}{{\bf{F}}_t^H}$. Furthermore, to make problem (\ref{eq44}) more tractable, we replace $\left\| {{{\bf{Y}}_D} - {{\bf{G}}^n}{f({{\bf b}})}} \right\|_F^2$ in the objective function with an upperbound surrogate function, which can be constructed by performing the same operations as the steps (c), (d), and (e) in (\ref{eq22}). As a result, we have
				\begin{equation}
					\begin{small}
						\begin{aligned}
							&\left\| {{{\bf{Y}}_D} - {{\bf{G}}^n}{f({{\bf b}})}} \right\|_F^2 \\
							=&{\rm{tr}}\left( {f{{({\bf{b}})}^H}{{\left( {{{\bf{G}}^n}} \right)}^H}{{\bf{G}}^n}f({\bf{b}})} \right) - 2{\rm{Re}} \left\{ {{\rm{tr}}\left( {f{{({\bf{b}})}^H}{{\left( {{{\bf{G}}^n}} \right)}^H}{{\bf{Y}}_D}} \right)} \right\} \\& + {\rm{tr}}\left( {{\bf{Y}}_D^H{{\bf{Y}}_D}} \right)\\
							\mathop  \le \limits^{(a)}&{\chi^{n-1}}\left\| f({\bf b}) \right\|_F^2 - 2{\mathop{\rm Re}\nolimits} \left\{ {{\rm tr}\left( {{f({\bf b})^H}}{{\bm \Xi}^{n-1}} \right)} \right\} + {\rm{constant}},\label{eq45}
						\end{aligned}
					\end{small}
				\end{equation}
				where ${{\bm \Xi}^{n-1}} = {\left( {\chi^{n-1}}{{\bf{I}}_{{N_t}}} - {\left({{{\bf{G}}^n}}\right)^H}{{{\bf{G}}^n}} \right)f({{\bf b}^{n-1}})} + {{{\left( {{{\bf{G}}^n}} \right)}^H}{{\bf{Y}}_D}}$, and ${\chi^{n-1}}$ is a parameter satisfying ${\chi^{n-1}}{{\bf I}_{N_t}} \succeq {\left({{{\bf{G}}^n}}\right)^H}{{{\bf{G}}^n}}$, which can be readily chosen as $\chi^{n-1} = \lambda_{\rm {max}}\left({{\left({{{\bf{G}}^n}}\right)^H}{{{\bf{G}}^n}}}\right)$. Consequently, the surrogate form of problem (\ref{eq44}) can be expressed as
				\begin{equation}
					%		\begin{small}
						\begin{aligned}	
							{{\bf{b}}^{n}} = \mathop { \arg \min }\limits_{{\bf{b}} \in {{[0,1]}^N}} {\chi ^{n - 1}}\left\| f({\bf{b}}) \right\|_F^2 - 2{\mathop{\rm Re}\nolimits} \left\{ {{\rm tr}\left( {{f({\bf{b}})}^H} {{\bm \Xi}^{n-1}}\right)} \right\} \\- \kappa \left\| {{\bf{b}} - {{\bf{0.5}}_N}} \right\|_2^2 + \frac{\rho }{2}{\left\| {\bf A}{\bf b}+ {{\bf{u}}^{n-1}} - {\bm \theta} + {{\bm \omega}^{n-1}} \right\|_2^2}, \label{eq46}
						\end{aligned}	
						%			\end{small}
				\end{equation}
				which takes almost the same form as problem (\ref{eq25}) and therefore the solution can be similarly achieved by		
				\begin{equation}
					\begin{aligned}
						b_{{i^{t,k,q}}}^n = \prod\nolimits_{\left[0,1\right]} {\left( {\frac{{\rho {\bf{a}}_{{i^{t,k,q}}}^T\left( {{\bm{\theta }} - {{\bf{u}}^{n - 1}} - {{\bm{\omega }}^{n - 1}}} \right) - {c^{n-1}_{{i^{t,k,q}}}} - \kappa }}{{\rho {\Lambda_{{i^{t,k,q}}}} + {e^{n-1}_{{i^{t,k,q}}}} - 2\kappa }}} \right)},\label{eq47}
					\end{aligned}
				\end{equation}
				where the expressions of ${e^{n-1}_{{i^{t,k,q}}}}$ and ${c^{n-1}_{{i^{t,k,q}}}}$ can be accordingly obtained like ${\beta^{n-1}_{{i^{t,k,q}}}}$ and ${\gamma^{n-1}_{{i^{t,k,q}}}}$ in (\ref{Ap1}) or (\ref{Ap2}), as long as we replace $\lambda ^{n-1}$ and $d_{kt}^{n-1}$ by $\chi ^{n-1}$ and $\xi_{kt}^{n-1}$, i.e., the $(k,t)$-th entry of ${\bm \Xi}^{n-1}$, respectively.   
				
				To be clear, we summarize the above JCDD procedure in Algorithm \ref{JCDD-S}. %, where line 14, analogous to line 13 in Algorithm \ref{JCDD-G}, also represents the early-stopping criterion.
				
				\begin{algorithm}[t]
					\small
					\caption{JCDD Receiving Algorithm for Sparse MmWave MIMO Channels}
					\begin{algorithmic}[1]
						\REQUIRE $\bf P$, ${\bf Y}_P$, ${\bf Y}_D$, $\bf{A}$, $\bm{\theta}$, $\bf{H}$, and $I_\text{max}$.%The received signals $ {{\bf{y}}}$, the imperfect CSI ${\hat{\bf h}}$, the  covariance matrix of CSI uncertainty ${{\bf{\Sigma }}_{\bf{h}}}$ and the maximum number of iteration $I_{max}$.
						\STATE Initialize: ${\bf{G}}_s^0$, ${\bf{b}}^{0}$, ${\bf{u}}^{0}$, ${\bm{\omega}}^{0}$, %$c^0$, 
						$n = 0$.
						\REPEAT 
						\STATE $n \gets n+1$.
						\STATE Update ${\bf{G}}_s^{n}$ via (\ref{eq40}) and calculate ${{\bf G}^n} = {{\bf{F}}_r}{{\bf G}_s^n}{{\bf{F}}_t^H}$.
						\FOR {$k=1,2,\cdots,{N_t},t=1,2,\cdots,{T_D}$}
						\STATE Sequentially update $b_{{i^{t,k,q}}}^n$ for $q=1,2,\cdots,Q$ via (\ref{eq47}).
						\ENDFOR
						\STATE Update ${\bf{u}}^{n}$ via (\ref{eq37b}).
						\STATE Update ${\bm{\omega }}^{n}$ via (\ref{eq37c}).
						%		\STATE ${c^{n}}\gets \frac{{1 + \sqrt {1 + 4{\left(c^{n - 1}\right)^2}} }}{2}$.
						%		\STATE ${\bf{u}}^{n} \gets {\bf{u}}^{n} + \frac{c^{n - 1}-1}{c^{n}}\left({\bf{u}}^{n} - {\bf{u}}^{n-1}\right)$.
						%		\STATE ${\bm{\omega}}^{n} \gets {\bm{\omega}}^{n} + \frac{c^{n - 1}-1}{c^{n}}\left({\bm{\omega}}^{n} - {\bm{\omega}}^{n-1}\right)$.
						\STATE ${\hat{b}}_i \gets 0$ if $b_i^{n}<0.5$, or ${\hat{ b}}_i \gets 1$ otherwise, for $i=1,2,\cdots,N$.
						\UNTIL{${\left[\sum\limits_{i = 1}^N {{H_{ji}}}{\hat{b}_i}\right]_2} = 0$ for $j=1,2,\cdots,M$ \textbf{or} $n = I_{\text{max}}$.}
						%\STATE Calculate $\hat{\bf G}_s$ with ${\hat{\bf b}}$ by iteratively performing (\ref{eq40}) until convergence.
						\ENSURE  ${\hat{\bf b}}$ and ${\hat{\bf{G}}} = {{\bf G}^n}$.
					\end{algorithmic}
					\label{JCDD-S}
				\end{algorithm}
				
				\subsection{Network Design}
\begin{figure}[!t]
			   					\centering
			   					\includegraphics[width=8cm]{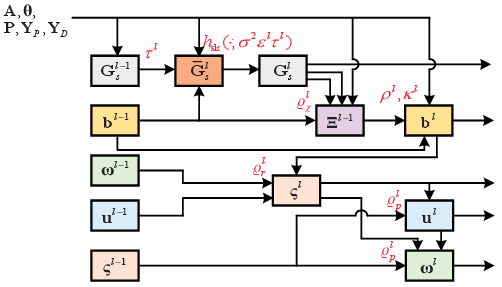}
			   					\caption{Block diagram of the $l$-th layer of JCDDNet-S.}
			   					\label{fig3}
			   				\end{figure}

				In the following, we develop another JCDD receiving network for sparse mmWave MIMO channels, referred to as JCDDNet-S. Akin to JCDDNet-G in Section~III-C, JCDDNet-S is built by unfolding the ADMM iterations of Algorithm 2 and utilizing the structure of the variant R-A-ADMM to improve the convergence. In the proposed network, the penalty parameters, the sparsity parameter, and the stepsize  involved in the ADMM iterations are set to be layer-wise trainable parameters, i.e., $\{{\rho}^l,{\kappa}^l,{\varepsilon}^l,{\tau}^l\}_{l = 1}^{L_\text{max}}$. Besides, we reexpress ${\chi^{n-1}}$ involved in (\ref{eq47}) as $ {\varrho_{\chi}^l}\lambda_{\rm {max}}\left({{\left({{{\bf{G}}^0}}\right)^H}{{{\bf{G}}^0}}}\right)$ so that only the factors $\{{\varrho_{\chi}^l}\}_{l = 1}^{L_\text{max}}$ need to be trained. Furthermore, we also add the trainable relaxation parameters $\{\varrho_r^l\}_{l = 1}^{L_\text{max}}$ and the predictor-corrector parameters $\{\varrho_p^l\}_{l = 1}^{L_\text{max}}$ to fit the R-A-ADMM structure. Thus, as shown in Fig. \ref{fig3} at the top of the next page, the $l$-th layer of the proposed network is formulated as follows:
			
			   	\begin{subequations}
			   	\label{eq5C1}
			   	\begin{equation}
			   		%				\begin{small}
			   			\begin{aligned}	
			   				{\bar{{\bf G}}_s^l} =  {{\bf G}_s^{l - 1}} - {\tau}^l \left( {{\bf{G}}_s^{l - 1}{\bf{F}}_t^H{\bf S}_{{\bf b}^{n-1}} - {\bf{F}}_r^H{\left[ {\bf Y} \right]}} \right){\left( {{\bf{F}}_t^H{\bf S}_{{\bf b}^{n-1}}} \right)^H}, \label{eq5C1_a}	
			   			\end{aligned}
			   			%	\end{small}
			   	\end{equation}
			   	\begin{equation}
			   		%			\begin{small}
			   			\begin{aligned}	
			   				{{\bf{G}}_s^l} = h_{\rm {sk}}\left({\bar{{\bf G}}_s^l};{\sigma ^2}{\varepsilon^l}{\tau^l}\right) , \label{eq5C1_b1}	
			   			\end{aligned}
			   			%	\end{small}		
			   	\end{equation}
			   	\begin{equation}
			   		%				\begin{small}
			   			\begin{aligned}	
			   				{{\bf{G}}^l} = {{\bf{F}}_r}{{\bf{G}}_s^l}{{\bf{F}}_t^H} , \label{eq5C1_b}
			   			\end{aligned}
			   			%\end{small}
			   		\end{equation}
			   		\begin{equation}
			   			%		\begin{small}
			   				\begin{aligned}	
			   					{{\bm \Xi}^{l-1}} = &{\left( {\varrho_{\chi}^{lr}}\lambda_{\rm {max}}\left({{\left({{{\bf{G}}^0}}\right)^H}{{{\bf{G}}^0}}}\right){{\bf{I}}_{{N_t}}} - {\left({{{\bf{G}}^l}}\right)^H}{{{\bf{G}}^l}} \right)f({{\bf b}^{l-1}})} \\ &+ {{{\left( {{{\bf{G}}^l}} \right)}^H}{{\bf{Y}}_D}}, \label{eq5C1_c}
			   				\end{aligned}
			   				%	 \end{small}
			   		\end{equation}
			   		\begin{equation}
			   			%		\begin{small}
			   				\begin{aligned}	
			   					{{\bf{b}}}^{l}\left[{\cal I}_q\right] =\prod\nolimits_{\left[0,1\right]^{N/Q}} \big( \big({\rho^l} {\bf{A}}\left[:,{\cal I}_q\right]^T\left( {{\bm{\theta }} - {{\bf{u}}^{l - 1}} - {{\bm{\omega }}^{l - 1}}} \right) \\ - {{\bf c}^{l-1}}\left[{\cal I}_q\right] - {\kappa^l}{{\bf{1}}_{N/Q}} \big) \oslash {\bm \zeta}^l\left[{\cal I}_q\right]\big), \forall q, \label{eq5C1_d}
			   				\end{aligned}
			   				%	\end{small}
			   		\end{equation}
			   		\begin{equation}
			   			%		\begin{small}
			   				\begin{aligned}	
			   					{\bf{b}}^{l} =  {\rm{Concat}}\left({\bf{b}}^{l}\left[{\cal I}_q\right],\forall q\right), \label{eq5C1_e}
			   				\end{aligned}
			   				%	\end{small}
			   		\end{equation}
			   		
			   		\begin{equation}
			   			%	\begin{small}
			   				\begin{aligned}	
			   					{\bm{\varsigma}}^{l} = {\bm \theta} - {\varrho_r^l}{\bf A}{{\bf b}^{l}}-\left(1-{\varrho_r^l}\right)\left({\bm \theta}-{\bf{u}}^{l-1}\right) - {\bm{\omega}}^{l-1}, \label{eq5C1_f}
			   				\end{aligned}
			   				%\end{small}
			   			\end{equation}
			   			\begin{equation}
			   				%	\begin{small}
			   					\begin{aligned}	
			   						{\bf{u}}^{l} = {\rm{ReLU}}\left({\bm{\varsigma}}^{l}\right) + {\varrho_p^l}\left({\rm{ReLU}}\left({\bm{\varsigma}}^{l}\right) - {\rm{ReLU}}\left({\bm{\varsigma}}^{l-1}\right)\right), \label{eq5C1_g} 
			   					\end{aligned}
			   					%\end{small}
			   				\end{equation}
			   				\begin{equation}
			   					%	\begin{small}
			   						\begin{aligned}	
			   							{\bm{\omega}}^{l} =  {\bf{u}}^{l} - \left(1+{\varrho_p^l}\right){\bm{\varsigma}}^{l} + {\varrho_p^l}{\bm{\varsigma}}^{l-1}, \label{eq5C1_h}
			   						\end{aligned}
			   						%\end{small}
			   					\end{equation}
			   				\end{subequations}
			   				where ${\bm \zeta}^l = {\rho^l}{{\rm diag}\left({{\bf A}^T}{\bf A}\right)} + {{\bf e}^{l-1}} - 2{\kappa^l}{{\bf{1}}_{N}}$ and ${\cal I}_q$ is defined in (\ref{eq4C5}) as ${\cal I}_q \triangleq \{ {i^{t,k,q}}\left| {\forall t,\forall k} \right.\}$. ${\bf e}^{l-1}$ and ${\bf c}^{l-1}$ involved in (\ref{eq5C1_d}) can be calculated  similarly to ${\bm \beta}^{l-1}$ and ${\bm \gamma}^{l-1}$  in (\ref{eq4C5_c}) by replacing ${\lambda}^{l-1}$ and ${{\bf S}^{l-1}}$ with ${\chi}^{l-1} = {\varrho_{\chi}^{l-1}}\lambda_{\rm {max}}\left({{\left({{{\bf{G}}^0}}\right)^H}{{{\bf{G}}^0}}}\right)$ and ${{\bm \Xi}^{l-1}}$, respectively.

									\section{Extension for Multiuser Case} 
Although the proposed two JCDD receivers for Gaussian MIMO channels and sparse mmWave MIMO channels are both derived for the single-user scenario, both of them can be easily extended to address the multiuser case. Specifically, we consider a $K$-user uplink MIMO system, where each user has $N_t$ transmit antennas. Let ${\bf H}_k \in {\left[0,1\right]}^{M_k \times N_k}$ denote the  parity-check matrix corresponding to the LDPC code adopted by the $k$-th user, where $M_k$ and $N_k$ represent the number of parity-check nodes and the code length, respectively. For the $k$-th user, the codeword ${\bf b}_k = \left[b_{k1},b_{k2},\cdots,b_{kN_k}\right]^T$ is mapped into $2^{Q_k}$-QAM symbols ${\bf x}_k = \left[x_{k1},x_{k2},\cdots,x_{k\left(N_k/Q_k\right)}\right]^T$ and then evenly allocated to each transmit antenna for spatial multiplexing transmission. For convenience, we still keep $N_k/{N_t}/Q_k = {T_D}$ for different users. The pilot matrix and data matrix become ${{\bf S}_P} \in \mathbb{C}^{K{N_t}\times{T_P}}$ and ${{\bf S}_D} = f(\underline{\bf b}) \in \mathbb{C}^{K{N_t}\times{T_D}}$, respectively, where $\underline{{\bf b}} =  \left[{\bf b}^T_1,{\bf b}^T_2,\cdots,{\bf b}^T_K\right]^T\in [0,1]^{\underline N}$ with ${\underline N} = {{\sum\limits_{k=1}^{K}{N_k}}}$,  and $f(\cdot)$ is the bit-to-symbol mapping function. Then, for Gaussian MIMO channels, by introducing the auxiliary variables ${\bf z}_k \in {\mathbb R}^{\Gamma_k^c}$ with $\Gamma_k^c = { \sum_{j=1}^{M_k}{{2^{\left| {{{\cal N}_c}(j)} \right| - 1}}}}$,  %${\bf z}_1, {\bf z}_2, \cdots, {\bf z}_K$, %matrix ${\bf Z}\in {\mathbb R}^{ \sum\limits_{j=1}^M{{2^{\left| {{{\cal N}_c}(j)} \right| - 1}}}\times K} $, 
problem (\ref{eq18}) can be updated by
\begin{equation} 
	\begin{aligned}
		\mathop {\min }\limits_{\underline{{\bf b}}\in [0,1]^{\underline N}}\quad&\Phi(\underline{{\bf b}})-\alpha \left\| {\underline{{\bf b}} - {\bf{0.5}}_{\underline N}} \right\|_2^2
		\\\text{s.t.}\quad\quad&{\bf A}_k{\bf b}_k+ {\bf z}_k = {\bm \theta}_k,{\bf z}_k\geq{\bf 0}, k = 1,2,\cdots,K, \label{eq56}
	\end{aligned}
\end{equation}
where $\Phi(\underline{{\bf b}})=-{{{\bf{y}}^H}{{\bf{X}}_{\underline{{\bf b}}}}{{\left( {{{\bf{X}}_{\underline{{\bf b}}}^H}{{\bf{X}}_{\underline{{\bf b}}}} + {\sigma ^2}{\bf{C}}_{\bf{g}}^{ - 1}} \right)}^{ - 1}}{{\bf{X}}_{\underline{{\bf b}}}^H}{\bf{y}}}$ and ${{\bf X}_{\underline{{\bf b}}}} \triangleq {\left[ {{\bf{S}}_P,f({\underline{{\bf b}}})} \right]^T}\otimes{{\bf I}_{N_r}}$. Note that ${\bf A}_k$ and ${\bm \theta}_k$ can be easily obtained according to ${\bf H}_k$ by following (\ref{eq12})-(\ref{eq14}). 
% \begin{equation} 
	%	\begin{aligned}
		%		\mathop {\min }\limits_{{\bf{B}} \in {\{0,1\} ^{N \times K}}}&\Phi({\bf B})-\alpha \left\| {{\bf{B}} - {\bf{0.5}}_{N \times K}} \right\|_2^2
		%		\\\text{s.t.}\quad\quad&{\bf A}{\bf B}+ {\bf Z}= {\bm \Theta},\quad{\bf U}\geq{\bf 0}, \label{eq56}
		%	\end{aligned}
	%\end{equation}
	%where $\Phi({\bf B})=-{{{\bf{y}}^H}{{\bf{X}}_{\bf B}}{{\left( {{{\bf{X}}_{\bf B}^H}{{\bf{X}}_{\bf B}} + {\sigma ^2}{\bf{C}}_{\bf{g}}^{ - 1}} \right)}^{ - 1}}{{\bf{X}}_{\bf B}^H}{\bf{y}}}$, ${{\bf X}_{\bf B}} \triangleq {\left[ {{\bf{P}},f({\bf B})} \right]^T}\otimes{{\bf I}_{N_r}}$, and ${\bm \Theta} = {\bm \theta} \otimes {\bf 1}_{1 \times K}$. 
	
	Similarly, for sparse mmWave MIMO channels, the multiuser extension of problem (\ref{eq35}) can be expressed by
	\begin{equation} 
		\begin{aligned}
			\mathop {\min }\limits_{{{\bf{G}}_s},{\underline{\bf{b}} \in {\{0,1\} ^{\underline N}}}} &\Psi({{\bf G}_s},\underline{{\bf b}}) -\kappa \left\| {\underline{{\bf b}} - {\bf{0.5}}_{\underline N}} \right\|_2^2\\
			\text{s.t.}\quad&{\bf A}_k{\bf b}_k+ {\bf u}_k = {\bm \theta}_k,{\bf u}_k\geq{\bf 0}, k = 1,2,\cdots,K,\label{eq57}
		\end{aligned}
	\end{equation}
	where $\Psi({{\bf G}_s},\underline{{\bf b}}) = \left\| {{\bf{F}}_r^H{\bf Y}- {{\bf{G}}_s}{\bf{F}}_t^H\left[ {{\bf{S}}_P,{f(\underline{{\bf b}})}} \right]} \right\|_F^2 + {\sigma ^2}\varepsilon {\left\| {{{\bf{G}}_s}} \right\|_1}$ and ${\bf u}_k$ denotes the auxiliary variable.
	%\begin{equation} 
	%	\begin{aligned}
		%		\mathop {\min }\limits_{{{\bf{G}}_s} \in {{\mathbb C}^{{N_r} \times {N_t}}},{{\bf{B}} \in {\{0,1\} ^{N \times K}}}} &\Psi({{\bf G}_s},{\bf B}) -\kappa \left\| {{\bf{B}} - {\bf{0.5}}_{N \times K}} \right\|_2^2\\
		%		\text{s.t.}\quad\quad\quad\quad&{\bf A}{\bf B}+ {\bf U}= {\bm \Theta},\quad{\bf U}\geq{\bf 0},\label{eq57}
		%	\end{aligned}
	%\end{equation}
	%where $\Psi({{\bf G}_s},{\bf B}) = \left\| {{\bf{F}}_r^H\left[ {{{\bf{Y}}_P},{{\bf{Y}}_D}} \right] - {{\bf{G}}_s}{\bf{F}}_t^H\left[ {{\bf{P}},{f({\bf B})}} \right]} \right\|_F^2 + {\sigma ^2}\varepsilon {\left\| {{{\bf{G}}_s}} \right\|_1}$ and ${\bf U}\in {\mathbb R}^{ \sum\limits_{j=1}^M{{2^{\left| {{{\cal N}_c}(j)} \right| - 1}}}\times K} $ is the auxiliary variable matrix.
	
	Algorithms 1 and 2 for problems (\ref{eq18}) and (\ref{eq35}) can be adopted to address problems (\ref{eq56}) and (\ref{eq57}) with minor modifications. Furthermore, the corresponding neural networks can also be readily designed by  adaptively revising previous steps. %The modifications are straightforward and thus not presented here due to the space limitation. 
	
	\section{Numerical Results}
	%This section evaluates the performances of the proposed JCDD networks in MIMO systems encoded by LDPC codes via numerical simulations. 
	
	%\footnote{5G LDPC codes can yield a better performance than the (3,6)-regular LDPC codes for the decoupled receivers, while the opposite behavior happens for both the turbo receivers and the proposed JCDD networks, which can be attributed to the noticeable proportion of of punctured bits in a short 5G codeword that can not benefit from the turbo iterations or the optimization of the objective function of the JCDD formulation. Therefore, the (3,6) regular codes are more preferable to achieve the high reliability required by URLLC.}
	
	%\subsection{Simulation Setup}
	In order to match the short packet transmission for URLLC, a short (3,6)-regular LDPC code ${\cal C}_1$ constructed by the progressive edge-growth (PEG) algorithm with code rate $R = 1/2$ and code length $N=288$ is used, and the number of pilot symbols is limited to $T_p = N_t$. For JCDDNet-G, we set $N_r = 8$ and $N_t = 4$. Both i.i.d. Gaussian channels ${\bf H}_{\text{i.i.d}}$ and spatially correlated Gaussian channels generated by the Kronecker model \cite{Kron5338}, i.e., ${\bf{H}}_{\text{cor}} = {\bf{R}}_r^{1/2}{{\bf{H}}_{{\rm{i}}{\rm{.i}}{\rm{.d}}}}{\bf{R}}_t^{1/2}$, are considered. Each entry of ${\bf H}_{\text{i.i.d}}$ follows a Gaussian distribution with zero mean and unit variance, ${{\bf R}_t}$ and ${{\bf R}_r}$ are the exponential correlation matrices of the transmitter and the receiver, which can be characterized by the correlation coefficients ${\rho_t}$ and ${\rho_r}$, respectively. { For JCDDNet-S, the Saleh-Valenzuela model is used to generate the mmWave MIMO channel, which can be characterized by $N_{cl}$ clusters with $N_p$ paths per cluster as follows \cite{mmModel}: 
		\begin{equation}
			{\bf{G}} = \sqrt{\frac{{N_t}{N_r}}{{N_{cl}}{N_p}}}\sum\limits_{c = 1}^{{N_{cl}}} {\sum\limits_{p = 1}^{{N_p}} {{\alpha _{c,p}}{{\bf{a}}_r}\left(\varphi _{c,p}^r,\vartheta _{c,p}^r\right){\bf{a}}_t^H\left(\varphi _{c,p}^t,\vartheta _{c,p}^t\right)} },\label{eq29}
		\end{equation}
		where ${\alpha _{c,p}}$, $\varphi_{c,p}^r(\vartheta_{c,p}^r)$, and $\varphi_{c,p}^t(\vartheta_{c,p}^t)$  denote the complex gain%distributed as ${\cal{CN}}\left(0,1\right)$
		, the azimuth (elevation) angle of arrival (AoA), and the azimuth (elevation) angle of departure (AoD) of the $p$-th path in the $c$-th cluster, respectively, and ${\bf a}_r$ and ${\bf a}_t$ are the receive and transmit array response vectors, respectively. In this work, we consider a half-wavelength spaced UPA of size ${N^y} \times {N^z}$, the array response vector of which can be expressed by
		\begin{equation}
			\begin{aligned}
				{\bf{a}}(\varphi,\vartheta) = \frac{1}{{\sqrt {{N^y}{N^z}}}}&\big[1,\cdots,{e^{\jmath\pi \left( {y\sin \varphi \sin \vartheta+z{\cos} {\vartheta}}\right)}},\cdots,\\&{e^{\jmath\pi \left( {({N^y} - 1)\sin \varphi \sin \vartheta  + ({N^z} - 1)\cos \vartheta } \right)}}\big]^T,\\&y=1,2,\cdots,{N^y},z=1,2,\cdots,{N^z},
			\end{aligned}
			\label{eq30}
		\end{equation}
		where the subscripts $r$ and $t$ are temporarily omitted for the simplicity of the expression. Specifically, we assume that the transmitter and the receiver are equipped with a $2 \times 2$ UPA and an $8 \times 8$ UPA, respectively, and there exist $N_{cl} = 8$ clusters with $N_p = 10$ paths per cluster.} The channel gains of the line-of-sight (LoS) paths and other paths are distributed as ${\cal{CN}}\left(0,1\right)$ and ${\cal{CN}}\left(0,10^{-0.5}\right)$, respectively. Besides, the azimuth and elevation AoAs and AoDs in each cluster are randomly generated with a mean cluster angle uniformly distributed over $[-\frac{\pi}{2},\frac{\pi}{2}]$ and the angular spread of 7.5 degrees.
	% and the angular spread of 7.5 degrees. The channel gains of the line-of-sight (LoS) paths and other paths are distributed as ${\cal{CN}}\left(0,1\right)$ and ${\cal{CN}}\left(0,10^{-0.5}\right)$, respectively. Besides, the azimuth and elevation AoA and AoD corresponding to the LOS path in each cluster are uniformly generated from $[-\frac{\pi}{2},\frac{\pi}{2}]$.
	
	\begin{figure*}[!t]
		
		\centering
		\subfigure[BLER performance]
		{\begin{minipage}[t]{0.49\textwidth}
				\centering
				\includegraphics[width=6.8cm]{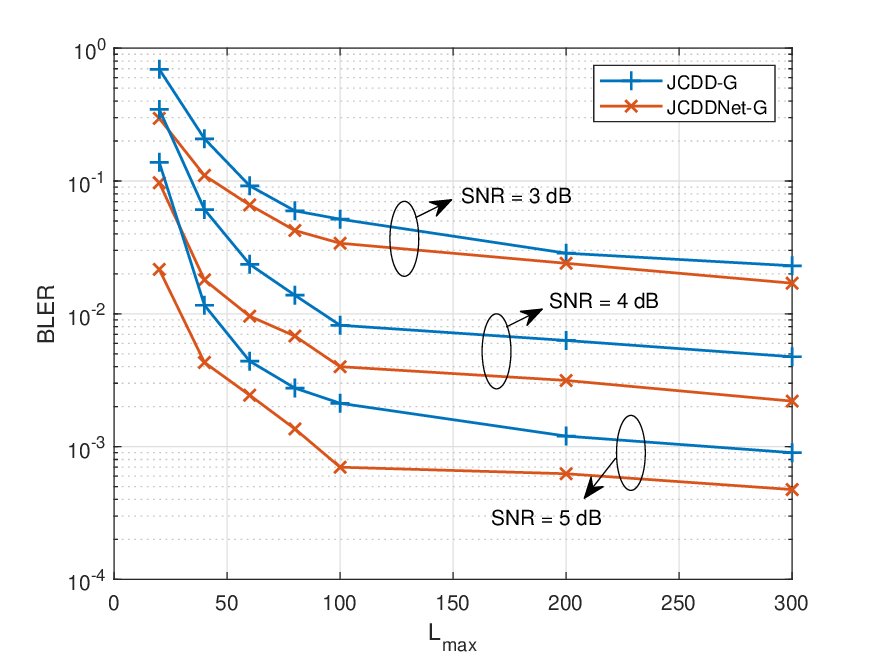}
			\end{minipage}		
		}		
		\subfigure[Average number of iterations/layers and time]
		{\begin{minipage}[t]{0.49\textwidth}
				\centering
				\includegraphics[width=6.8cm]{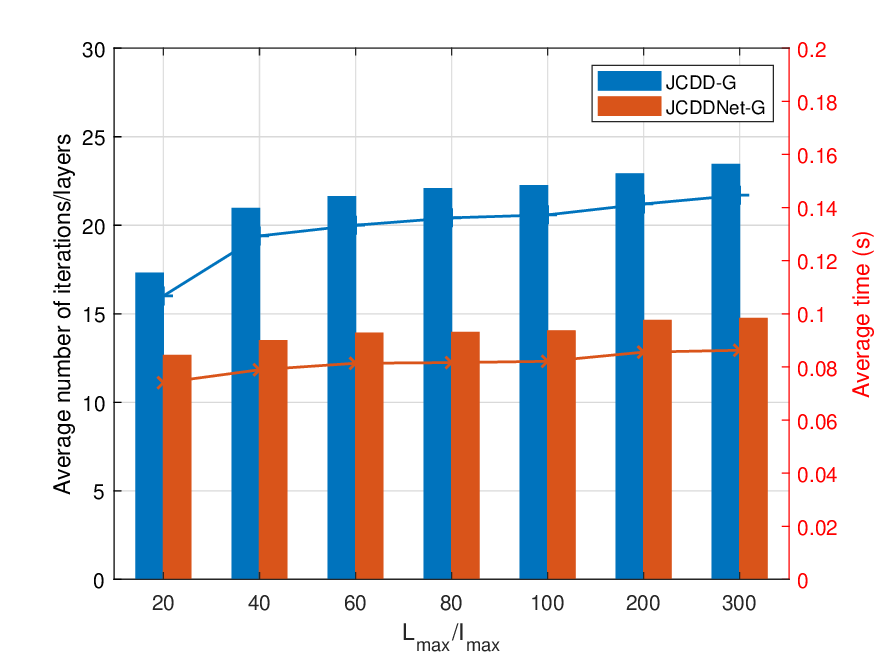}
			\end{minipage}		
		}
		%			\subfigure[Average Time]
		%			{\begin{minipage}[t]{0.3\textwidth}
				%				\centering
				%				\includegraphics[width=6cm]{time.eps}
				%			\end{minipage}		
			%			}	
		\caption{Comparison of JCDDNet-G and its corresponding ADMM algorithm in terms of (a) BLER performance and (b) average number of iterations/layers and time per codeword under i.i.d. Gaussian MIMO channels with QPSK modulation.}
		\label{fig4}
	\end{figure*}
	The traditional decoupled and turbo receivers, including the IDD \cite{Achieving2003,ASIC7894280} and the ICDD \cite{MICED}, are adopted for performance comparison, where the minimum mean squared error (MMSE), {the generalized approximate message passing (GAMP) \cite{GAMP},} and the maximum a posterior probability (MAP)-based MIMO detectors are taken into account for each kind of receiver. Specifically, the linear MMSE (LMMSE)-based channel estimation method is used for Gaussian MIMO channels, while the iterative shrinkage thresholding algorithm (ISTA) \cite{ISTA} is used to deal with the convex Lasso channel estimation problem for sparse mmWave MIMO channels. For simplicity, we denote these baselines as {MMSE/MAP-Decouple}, {MMSE/GAMP/MAP-IDD}, and {MMSE/GAMP/MAP-ICDD}. {Specifically, the channel estimation MSE-based virtual pilot selection method proposed in \cite{SPCreceiver} is also considered as a baseline for the Gaussian channels and denoted by MMSE-ICDD (Selected), where the number of the selected symbols $T_V$ are set to $\frac{3}{4}T_D$ and $\frac{2}{3}T_D$  for the QPSK and the 16QAM cases, respectively.} Moreover, the maximum number of GAMP iterations, the maximum number of BP iterations for LDPC decoding, the maximum number of turbo iterations, and the maximum number of ISTA iterations are set to $10$, $100$, $10$ and $100$, respectively. Note that the early termination mechanism is also applied in these baseline receivers\footnote{Source code: https://github.com/sunyi0101/JCDDNet}.
	
	\newcounter{TempEqCnt4}
	\setcounter{TempEqCnt4}{\value{equation}} 
	\setcounter{equation}{51} 	
	\begin{figure*}[hb] 
		\hrulefill
		% 		 		\begin{small}
			\begin{align}		
				{\cal L}\left(\{{\bm \Upsilon}_l\}_{l = \left(s-1\right)L_{\text{part}}+1}^{sL_{\text{part}}}\right)  = \frac{1}{{\left| {\cal T} \right|}}\sum\limits_{\cal T} \sum\limits^{sL_{\text{part}}}_{l=\left(s-1\right)L_{\text{part}}+1}{\left\| { \tanh\left(\varsigma\left({\bf{b}}^{l}- {\bf {0.5}}_N\right)\right) - \left(2{{\bf{b}}}- {\bf 1}_N\right) } \right\|_2^2}, \label{eq_loss}
			\end{align}
		\end{figure*} 
		\setcounter{equation}{\value{TempEqCnt4}} 
		
		\subsection{Training Strategy}

		\begin{figure*}[!t]
			
			\centering
			\subfigure[i.i.d. Gaussian channel]
			{\begin{minipage}[t]{0.49\textwidth}
					\centering
					\includegraphics[width=6.8cm]{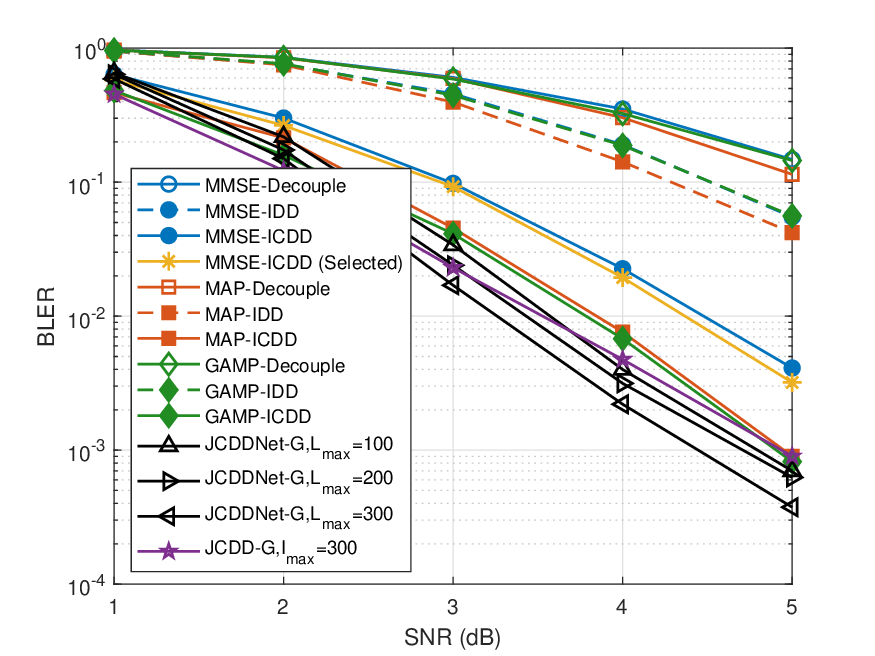}
				\end{minipage}		
			}
			\subfigure[Correlated Gaussian channel]
			{\begin{minipage}[t]{0.49\textwidth}
					\centering
					\includegraphics[width=6.8cm]{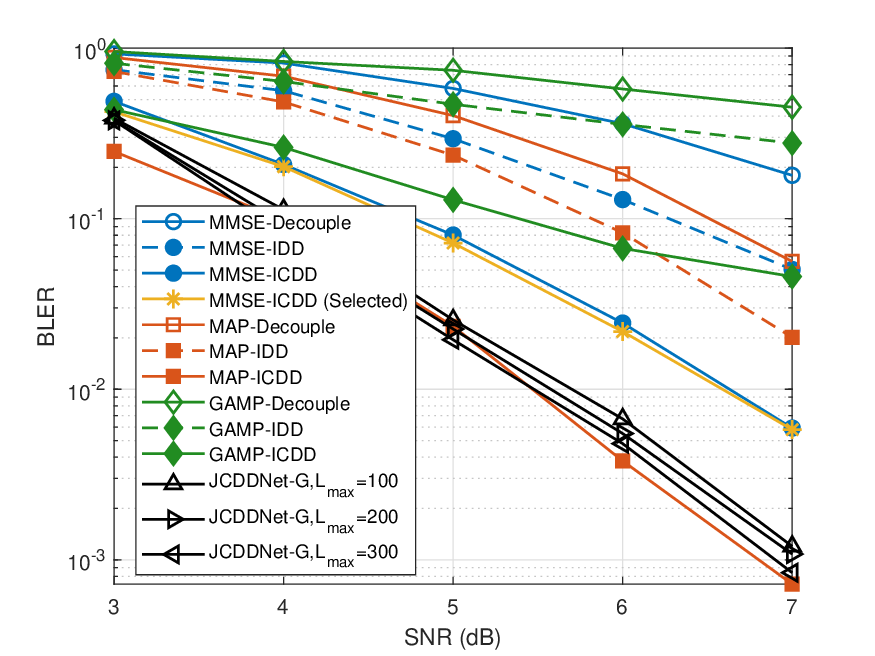}
				\end{minipage}		
			}	
			\caption{BLER performances of different receivers under (a) i.i.d. and (b) correlated Gaussian channels with QPSK modulation.}
			\label{fig5}
		\end{figure*}
		\begin{figure*}[!t]
			
			\centering
			\subfigure[i.i.d. Gaussian channel]
			{\begin{minipage}[t]{0.49\textwidth}
					\centering
					\includegraphics[width=6.8cm]{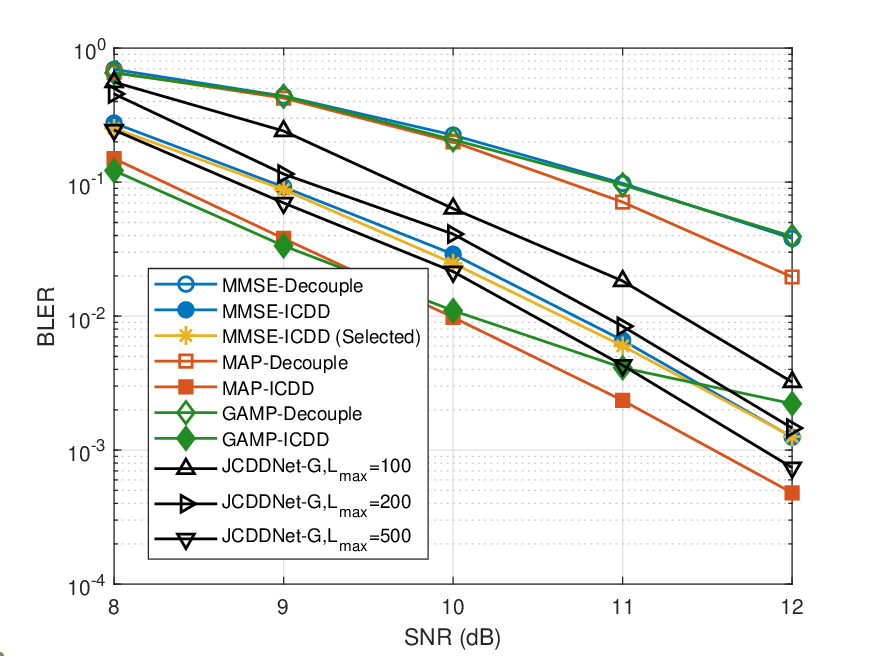}
				\end{minipage}		
			}
			\subfigure[Correlated Gaussian channel]
			{\begin{minipage}[t]{0.49\textwidth}
					\centering
					\includegraphics[width=6.8cm]{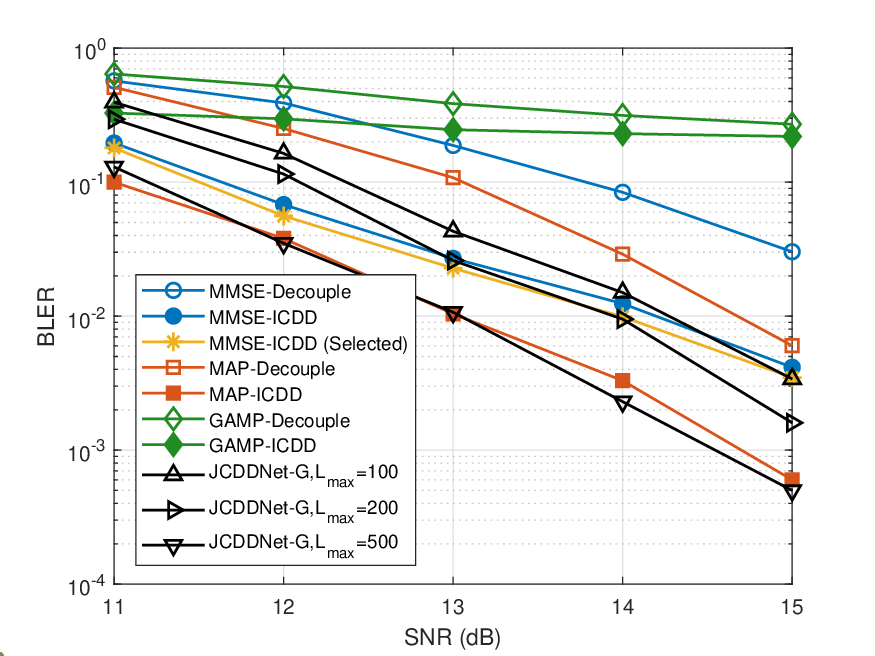}
				\end{minipage}		
			}	
			\caption{BLER performances of different receivers under (a) i.i.d. and (b) correlated Gaussian channels with 16QAM modulation.}
			\label{fig6}
		\end{figure*}
		
		%Once the training is finished, the proposed networks with the optimized parameters can be directly regarded as iterative algorithms for online prediction, where the early termination mechanism, as a common practice in LDPC-coded systems, can be adopted to save the computational overhead. 
		\vspace{-1mm}	
		The proposed networks are trained offline with 10000 training samples under a specific SNR point where the block error rate (BLER) of the original ADMM algorithm reaches 1e-2, which can provide enough error patterns to learn the underlying decoding philosophy. Then, we test the networks under different SNRs using continuously generated samples until 100 error blocks are collected. {Note that the training and the validation datasets are generated under different channel realizations with a consistent channel statistical distribution. After the offline training, the learned networks can be directly used for a new channel realization and there is no need for retraining as long as the channel statistics remains unchanged.} However, the early-termination mechanism can not be easily incorporated into the training process since the number of network layers should be fixed. %while 
		On the other hand, although we can predefine a sufficiently large number of layers to ensure the convergence, the training of such a large-scale network is quite challenging. Specifically, the traditional end-to-end training can suffer from a high training complexity and the vanishing gradient problem, and the layer-by-layer training can lead to a noticeable performance loss.

		To address the training issue, we present a multi-stage multi-layer training strategy, which means dividing the network training into several stages. For the $s$-th stage, we only need to train the current $L_{\text{part}}$ layers with the parameters of the preceding $\left(s-1\right)L_{\text{part}}$ layers frozen. By taking all the layers into account, we define the loss function in (\ref{eq_loss}) at the bottom of the page,
		%		\begin{equation}
			%			\begin{small} 
				%			\begin{aligned}
					%				&{\cal L}\left(\{{\bm \Upsilon}_l\}_{l = \left(s-1\right)L_{\text{part}}+1}^{sL_{\text{part}}}\right)  = \\ &\frac{1}{{\left| {\cal T} \right|}}\sum\limits_{\cal T} \sum\limits^{sL_{\text{part}}}_{l=\left(s-1\right)L_{\text{part}}+1}{\left\| { \tanh\left(\varsigma\left({\bf{b}}^{l}- {\bf {0.5}}_N\right)\right) - \left(2{{\bf{b}}}- {\bf 1}_N\right) } \right\|_2^2}, \label{eq_loss}
					%			\end{aligned}
				%				\end{small}
			%		\end{equation}
		where ${\bm \Upsilon}_l$ denotes the trainable parameters of the $l$-th layer, $\cal T$ is the set of training samples,  and $\tanh\left(\varsigma\left(\cdot\right)\right)$ with $\varsigma$ being a large number, e.g., 200, is  used to approximate the sign function while promising the differentiability. By minimizing (\ref{eq_loss}), the output of each layer is considered to approach the labels, which is friendly to the early termination mechanism in the online phase. According to the simulation results, We fix the number of network layers to $100$ for an efficient training%\footnote{We have also conducted the experiments to train more layers, which results in limited benefits.}
		. Note that it is allowed to set $L_{\text{max}}$ to be larger than $100$ for online prediction, where the parameters of the networks are loaded with the well-trained values in the first 100 layers and remain to be the default values\footnote{We use a coarse grid search to optimize the penalty parameters involved in the ADMM algorithms, which are also regarded as the default values of the corresponding parameters in the network. On the other hand, for the additional introduced parameters in the network, the default values are set as $o_r^l = 1$, $o_p^l = 0$, $o_{\lambda}^l = 1$, and $o_{\upsilon}^l = 1$ for each layer of JCDDNet-G. The default values of the parameters of JCDDNet-S are determined in a similar way.} in the subsequent layers. In addition, we empirically set $L_{\text{part}} = 20$ for a reasonable tradeoff between the training complexity and the overall performance. The Adam optimizer is applied, where the batchsize, the number of epochs, and the learning rate are set to 200, 100, and 0.01 for each stage, respectively.
		
		\subsection{Performance Evaluation}
		We first compare the proposed JCDDNet-G with its corresponding ADMM algorithm to verify the effectiveness of the training. Fig. \ref{fig4}(a) shows their BLER performances in each stage under i.i.d Gaussian MIMO channels with QPSK modulation, where we also provide the results with $L_{\text{max}}=200$ and $L_{\text{max}}=300$ for reference. It can be seen that, benefiting from the well-trained parameters, JCDDNet-G can always outperform the original ADMM algorithm JCDD-G and the performance advantage is more obvious for a higher SNR point. On the other hand, we also count the average number of iterations/layers and time required per codeword of the two considered receivers when using the early termination mechanism, and the results at SNR = 4 dB are provided for an intuitive illustration in Fig. \ref{fig4}(b). We can observe that, JCDDNet-G only requires an average of around 13 iteration, which is much less than the counterpart, indicating a faster convergence rate. {The corresponding average time also shows a consistent behavior.} Meanwhile, increasing $L_{\text{max}}$ to $300$ can bring limited performance gains and negligible complexity burdens, since only a very few tasks need hundreds of iterations. A similar trend can be found for JCDDNet-S and its corresponding ADMM algorithm, which is not presented due to the space limitation.
		
		\subsubsection{Gaussian MIMO channels} We now investigate the performances of the proposed JCDDNet-G and turbo receivers under i.i.d and correlated Gaussian channels with  ${\rho_t}={\rho_r} = 0.5$, respectively. Fig. \ref{fig5} compares the performances of different receivers under QPSK modulation. As shown in Fig. \ref{fig5}(a), the IDD receivers can obtain a performance gain over the decoupled receivers by iteratively exchanging information between the detector and the decoder. However, the gain is relatively limited owing to the coarse CSI acquired based on the few pilot symbols, leading to severe error propagation. As an improved alternative, the ICDD receivers utilize the decoding output to generate virtual pilots to aid the channel estimation, which can therefore yield a much better performance than IDD receivers. { Furthermore, the selection of virtual pilots can only bring limited performance gains while introducing additional complexities of calculating the MSE metric for each soft data symbol and sorting, making the method less cost-efficient.} On the other hand, by directly integrating the functions of different modules and optimizing the involved parameters via offline training, the proposed JCDDNet-G is immune to the error propagation and also endowed with robustness. Consequently, it can considerably outperform {MAP-ICDD}. { In fact, we mention that even without the deep learning-based optimization, JCDD-G with the fixed parameters determined based on a coarse grid search method can also outperform the turbo receivers, suggesting the superiority of the proposed JCDD framework. It is also interesting to note that the GAMP-ICDD can sometimes perform slightly better than the MAP-ICDD, since the local optimality of sub-modules does not necessarily guarantee global optimality. Moreover, the information exchanged in turbo iterative algorithms relies heavily on empirical principles, which might suffer from some uncertainties in theory, whereas our proposed JCDD algorithms are based on theoretical formulations.} 	In Fig. \ref{fig5}(b), the gains achieved by the turbo receivers under correlated Gaussian MIMO channels are similar to those in Fig. \ref{fig5}(a), {while the GAMP-based receivers derived for i.i.d. Gaussian channels suffer from dramatic performance degradation in presence of channel correlation.} Fortunately, JCDDNet-G can still maintain a comparable performance to {MAP-ICDD}.

		\begin{figure*}[!t]
			
			\centering
			\subfigure[QPSK]
			{\begin{minipage}[t]{0.49\textwidth}
					\centering
					\includegraphics[width=6.8cm]{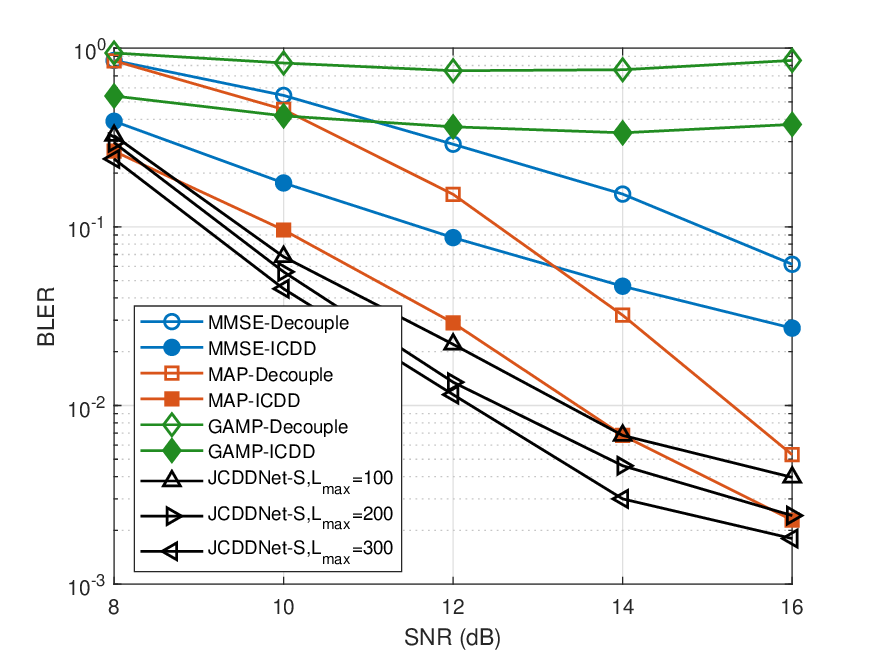}
				\end{minipage}		
			}
			\subfigure[16QAM]
			{\begin{minipage}[t]{0.49\textwidth}
					\centering
					\includegraphics[width=6.8cm]{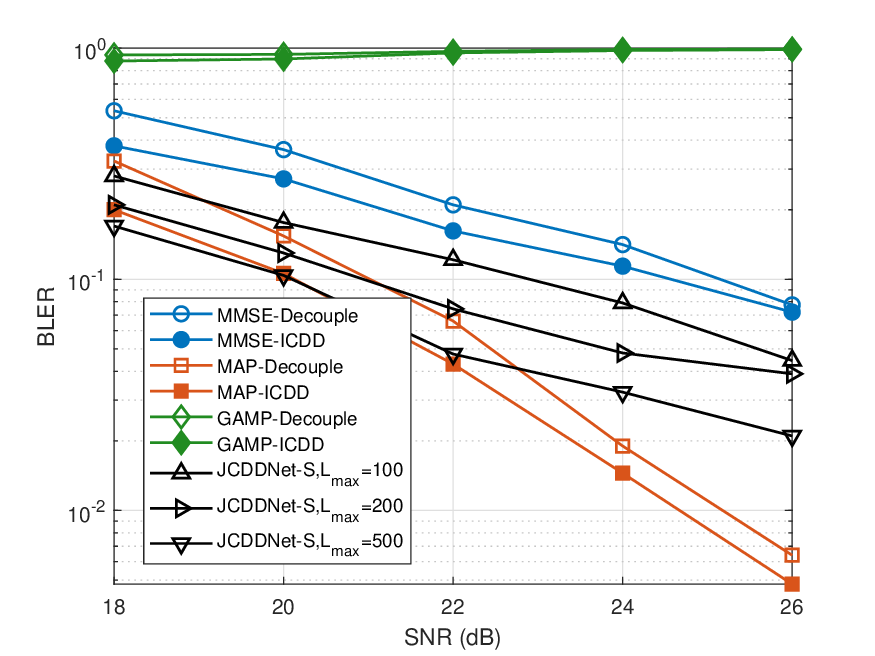}
				\end{minipage}		
			}	
			\caption{BLER performances of different receivers under sparse mmWave channels with different modulation orders.}
			\label{fig7}
		\end{figure*}
		\begin{figure*}[!t]
			\centering
			\subfigure[JCDDNet-G]
			{\begin{minipage}[t]{0.49\textwidth}
					\centering
					\includegraphics[width=6.8cm]{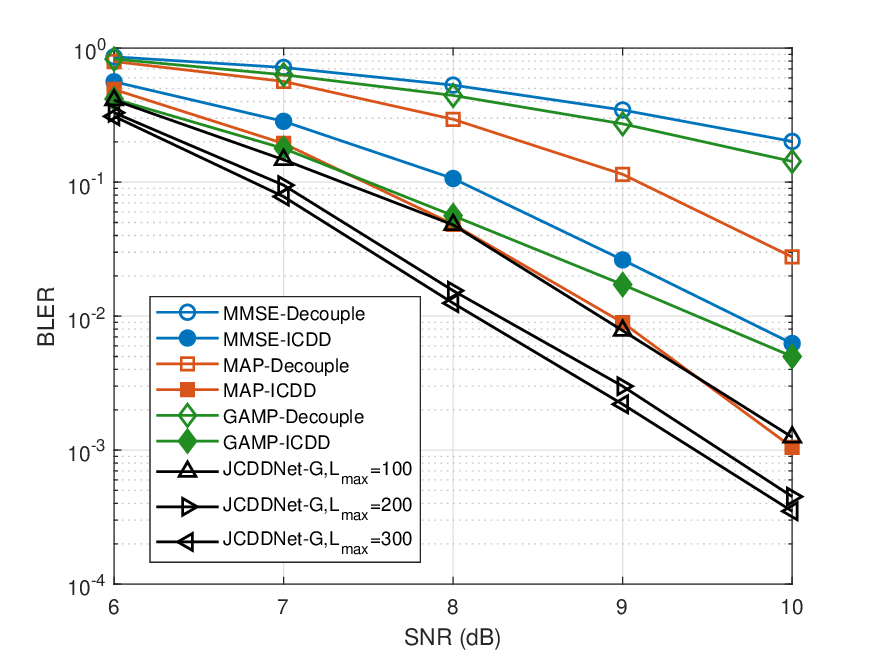}
				\end{minipage}		
			}
			\subfigure[JCDDNet-S]
			{\begin{minipage}[t]{0.49\textwidth}
					\centering
					\includegraphics[width=6.8cm]{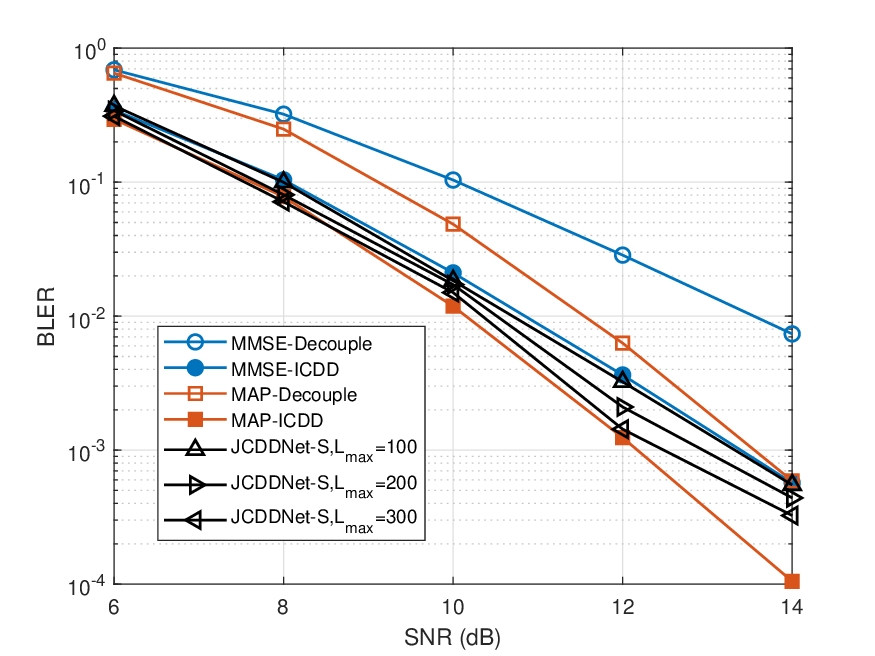}
				\end{minipage}		
			}	
			\caption{BLER performances of (a) JCDDNet-G and (b) JCDDNet-S for a 2-user system using code ${\cal C}_2$ and QPSK modulation.}
			\label{fig8}
		\end{figure*}				
		The performances of JCDDNet-G under Gaussian MIMO channels with 16QAM modulation are illustrated in Fig. \ref{fig6}. Concerning the inferior performances of IDD receivers as discussed above, we do not plot their results for the simplicity of the curves. We can see that, JCDDNet-G performs better than the decoupled receivers with $L_{\text{max}}=100$ and can approach the high-complexity {MAP-ICDD} with $L_{\text{max}}=500$. It is also noteworthy that the performance improvement by increasing $L_{\text{max}}$ is more obvious than that in the QPSK case, especially for correlated Gaussian channels considered in Fig. \ref{fig6}(b), demonstrating a slower convergence rate for a higher-order modulation. This can be attributed to the fact that the solution (\ref{eq28}) to subproblem (\ref{eq26}) is optimal in the QPSK case but suboptimal in the 16QAM case. 
		
		\vspace{-1mm}																										\subsubsection{Sparse mmWave MIMO channels} The performances of the other designed network JCDDNet-S under sparse mmWave MIMO channels with different modulation orders are given in Fig. \ref{fig7}. {Due to the partially known channel distribution and the low-rank and highly correlated nature of mmWave MIMO channels, it can be found that all the considered receivers suffer from slowly descending BLER curves versus SNR (Note that the SNR interval between two plotted points is 2 dB here), where the GAMP-based receivers can even hardly work.} Meanwhile, the turbo gains obtained by the ICDD receivers compared to the decoupled receivers also become limited in the 16QAM case as a result of the severe error propagation. As illustrated in Fig. \ref{fig7}(a), our proposed JCDDNet-S can always remarkably outperform {MMSE-ICDD} and approach {MAP-ICDD} with $L_{\text{max}} = 100$ in the QPSK case. However, for 16QAM considered in Fig. \ref{fig7}(b), there is still an evident gap from the MAP-based receivers even when $L_{\text{max}}$ is increased to $500$. Nevertheless, JCDDNet-S enjoys a much lower complexity, as will be analyzed in the next subsection.

		\vspace{-1mm}																											\subsubsection{Multiuser case} This part provides some results to validate the adaptability of the proposed networks to multiuser cases. Note that the network scale is directly related to the total number of bit variables, which can greatly influence the required computational resources and training time. For a convenient verification, we only consider a small-scale scenario with $K = 2$ users, where each $N_t = 4$-antenna user uses a short (3,6)-code ${\cal C}_2$ with code rate $R = 1/2$ and code length $N = 144$ and adopts the QPSK modulation. 
		Fig. \ref{fig8}(a) presents the performances of JCDDNet-G for the considered multiuser scenario. We can observe that, JCDDNet-G can not reach convergence with 100 layers as in the single-user scenario, which suggests that it requires more layers to decode more data streams. Moreover, a 0.5 dB performance gain beyond {MAP-ICDD} can be attained with $L_{\text{max}} = 200$.  
		
		\begin{figure}[!t]
			
			\centering
			\includegraphics[width=6.8cm]{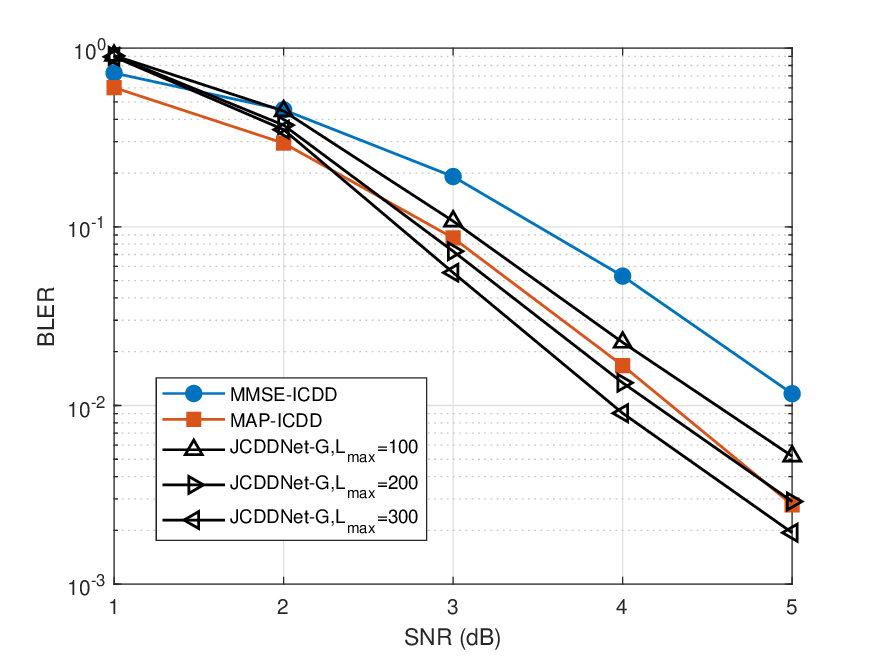}
			\caption{BLER performance of JCDDNet-G under correlated Gaussian channels with $\rho_t = \rho_r = 0.3$ by using the parameters trained under i.i.d. Gaussian channels.}
			\label{fig9}
		\end{figure}
		
		\begin{figure}[!t]
			
			\centering
			\includegraphics[width=6.8cm]{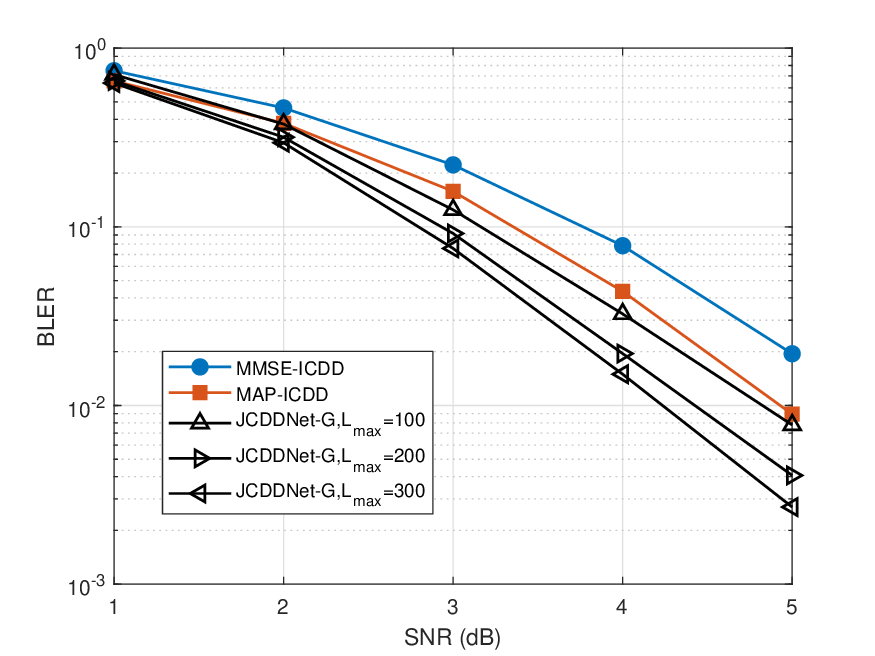}
			\caption{BLER performance of JCDDNet-G under i.i.d Gaussian channels with QPSK modulation and ${\cal C}_2$ of codelength $N = 144$ by using the parameters trained with ${\cal C}_1$ of codelength $N = 288$.}
			\label{fig10}
		\end{figure}
		
		On the other hand, we mention that the low-rank mmWave MIMO channels can not support a reliable transmission of full data streams in the multiuser case. Therefore, we reduce the number of streams for each user to $N_s = 2$ and use a precoding matrix ${\bf W} \in {\mathbb{C}}^{{N_t} \times {N_s}}$ to map the streams onto $N_t = 4$ transmit antennas. In the simulation, we fix ${\bf W}={\frac{1}{2}}{\left[1,0;0,1;1,0;0,\jmath\right]}$, which is selected from \cite{3GPP}. The corresponding results are plotted in Fig. \ref{fig8}(b). Thanks to the array gain provided by $\bf W$, the considered receivers can behave well in the multiuser mmWave scenario, where JCDDNet-S can still show a satisfying performance that is better than {MMSE-ICDD} and also comparable to {MAP-ICDD}.  
		%\begin{array}{cccc} 1&0&1&0 \\ 0&1&0& \jmath \end{array}

		{\subsubsection{Generalization ability}  Note that the proposed JCDD networks are model-driven and developed by unfolding the ADMM iterations of the derived receiving algorithms, which only involve a few trainable parameters and can be naturally less prone to overfitting. The previous results have verified the generalization ability of the proposed JCDD network under mismatched SNRs. Furthermore, we conduct simulations to test the robustness against mismatched channel distributions and mismatched codelengths. Fig. \ref{fig9} presents the performance of JCDDNet-G by directly applying the parameters trained under i.i.d. Gaussian channels for correlated Gaussian channels $\rho_t = \rho_r = 0.3$. It can be observed that the mismatched network exhibits robustness to the low correlated Gaussian channels and can still outperform the MMSE-based and MAP-based turbo receivers. Additionally, in Fig. \ref{fig10}, JCDDNet-G is directly loaded with the parameters trained for ${\cal C}_1$ of codelength $N = 288$ and evaluated with ${\cal C}_2$ of codelength $N = 144$, where the superiority over {MAP-ICDD} is also maintained. We also admit that the proposed networks fail to generalize for a significantly deviated channel distribution or a code with mismatched distributions of variable degrees. Fortunately, it is feasible to train the networks for different spatial correlation levels or different codes in advance and then pre-store look-up tables for online applications.}
		
		\begin{figure}[!t]
			\centering
			\includegraphics[width=6.8cm]{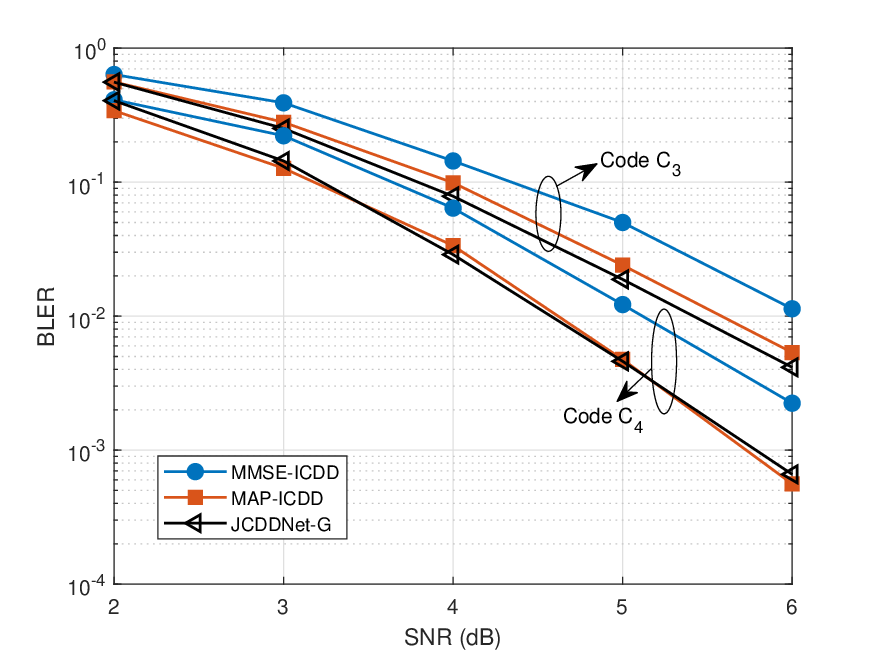}
			\caption{BLER performances of different receivers under i.i.d. Gaussian channels with QPSK modulation using CCSDS codes ${\cal C}_3$ and ${\cal C}_4$.}
			\label{fig12}
		\end{figure}
		
		{ \subsubsection{Other codes}We also test the proposed JCDD receivers using the commonly used consultative committee for space data systems (CCSDS) codes, i.e., ${\cal C}_3$ of codelength $N = 128$ and code rate 1/2, and ${\cal C}_4$ of codelength $N  = 256$ and code rate 1/2 \cite{channelcodes}. It can be seen from Fig. \ref{fig12} that, the proposed JCDDNet-G can still obviously outperform  MMSE-ICDD, while the performance advantage over MAP-ICDD becomes limited when the codelength increases from 128 to 256. Similar trends can be also found for 5G codes, which are not presented due to the strict page limit. }

		\begin{table*}[!t]
			\centering
			\caption{Complexity per Codeword of Different Modules/Receivers}
			{%
				\begin{tabular}{|ccc|cc|}
					\hline
					\multicolumn{3}{|c|}{\textbf{Modules/Receivers}}                        & \multicolumn{2}{c|}{\textbf{Complexity}} \\ \hline
					\multicolumn{1}{|c|}{\multirow{3}{*}{\makecell[c]{Channel \\ Estimation}}} & \multicolumn{1}{c|}{\multirow{2}{*}{LMMSE}} & {\makecell[c]{i.i.d.\\Gaussian}} & \multicolumn{1}{c|}{\makecell[c]{$C^{\text{Ini}}_{\text{ChE}} = 8{N_t^3}+8{N_t^2}+3{N_t}$\\$+16{N^2_t}{T_P}+8{N_r}{N_t}{T_P}$}} &  {\makecell[c]{$C^{\text{Iter}}_{\text{ChE}} = 4{N_t}Q{T_D}+(Q+4){N_t}2^Q{T_D}+ 8{N_t^3}$\\$+8{N_t^2}+3{N_t}+16{N^2_t}T+8{N_r}{N_t}T$}} \\ \cline{3-5} 
					\multicolumn{1}{|c|}{}         & \multicolumn{1}{c|}{}     & {\makecell[c]{Correlated\\Gaussian}} & \multicolumn{1}{c|}{\makecell[c]{$C^{\text{Ini}}_{\text{ChE}} = 8{N_r^3}{N_t^3}+10{N^2_r}{N^2_t}+2{N_r}{N_t}$\\$+8{N_r^3}{N^2_t}{T_P}+8{N^2_r}{N_t}{T_P}+8{N^2_t}{T_P}$}}         &   {\makecell[c]{$C^{\text{Iter}}_{\text{ChE}} = 4{N_t}Q{T_D}+(Q+4){N_t}2^Q{T_D}$\\$+ 8{N_r^3}{N_t^3}+10{N^2_r}{N^2_t}+2{N_r}{N_t}$\\$+8{N_r^3}{N^2_t}{T}+8{N^2_r}{N_t}{T}+8{N^2_t}{T}$}}       \\ \cline{2-5} 
					\multicolumn{1}{|c|}{}         & \multicolumn{1}{c|}{ISTA} & {\makecell[c]{Sparse\\mmWave}}     & \multicolumn{1}{c|}{\makecell[c]{$C^{\text{Ini}}_{\text{ChE}} =I_{\text{ISTA}}\left(8{N_r}{N^2_t}+10{N_r}{N_t}\right)$\\$+16{N_t^2}{T_P}+8{N_r^2}{T_P}+8{N_r}{N_t}{T_P}$\\$+8{N_r^2}{N_t}+8{N_r}{N^2_t}$}}         &    {\makecell[c]{$C^{\text{Iter}}_{\text{ChE}} =4{N_t}Q{T_D}+(Q+4){N_t}2^Q{T_D}$\\$+I_{\text{ISTA}}\left(8{N_r}{N^2_t}+10{N_r}{N_t}\right)+16{N_t^2}{T}$\\$+8{N_r^2}{T}+8{N_r}{N_t}{T}+8{N_r^2}{N_t}+8{N_r}{N^2_t}$}}      \\ \hline
					\multicolumn{1}{|c|}{\multirow{3}{*}{\makecell[c]{Data \\ Detection}}}     & \multicolumn{2}{c|}{MMSE}                            & \multicolumn{1}{c|}{\makecell[c]{$C^{\text{Ini}}_{\text{Det}} = 8{N_t^3}+8{N_t^2}+24{N^2_r}{N_t}$\\$+5{N_t}+8{N_r}{N_t}{T_D}+7{N_t}2^Q{T_D}$\\$+\left(2^Q+1\right){N_t}Q{T_D}$}} & {\makecell[c]{$C^{\text{Iter}}_{\text{Det}} =  (Q+7){N_t}2^Q{T_D}+8{N_r^3}{T_D}+8{N_r^2}{T_D}$\\$+2{N_r}{T_D}+16{N^2_r}{N_t}T_D+8{N_r}{N^2_t}T_D$\\$+16{N_r}{N_t}T_D+4{N_r}{T_D}+\left(2^Q+1\right){N_t}Q{T_D}$}} \\ \cline{2-5} 
					\multicolumn{1}{|c|}{}         & \multicolumn{2}{c|}{GAMP}              & \multicolumn{1}{c|}{\makecell[c]{$C^{\text{Ini}}_{\text{Det}} = I_{\text{GAMP}}\big(35{N_r}{N_t}{T_D}+8{N_r}{T_D}$\\$+7{N_t}{T_D}+15{N_t}2^Q{T_D}\big)$\\$+\left(2^Q+1\right){N_t}Q{T_D}$}}         &     {\makecell[c]{$C^{\text{Iter}}_{\text{Det}} =I_{\text{GAMP}}\big(35{N_r}{N_t}{T_D}+8{N_r}{T_D}$\\$+7{N_t}{T_D}+\left(15+Q\right){N_t}2^Q{T_D}\big)$\\$+Q{N_t}2^Q{T_D}+\left(2^Q+1\right){N_t}Q{T_D}$}}     \\ \cline{2-5} 
					\multicolumn{1}{|c|}{}         & \multicolumn{2}{c|}{MAP}               & \multicolumn{1}{c|}{\makecell[c]{$C^{\text{Ini}}_{\text{Det}} = \left(6{N_r}+8{N_r}{N_t}\right)2^{{N_t}Q}{T_D}$\\$+\left(2^{{N_t}Q}+1\right){N_t}Q{T_D}$}}         &     {\makecell[c]{$C^{\text{Iter}}_{\text{Det}} =\left(6{N_r}+8{N_r}{N_t}-1\right)2^{{N_t}Q}{T_D}$\\$+\left(2^{{N_t}Q+1}+1\right){N_t}Q{T_D}$}}     \\ \hline
					\multicolumn{1}{|c|}{Decoding} & \multicolumn{2}{c|}{BP}                & \multicolumn{2}{c|}{$C_{\text{Dec}} = {I_{\text{BP}}}\left(4M{d_c} + M{d^2_c}\right)$}                    \\ \hline
					\multicolumn{3}{|c|}{Decoupled Receiver}                                         & \multicolumn{2}{c|}{$C^{\text{Ini}}_{\text{ChE}} + C^{\text{Ini}}_{\text{Det}} + C_{\text{Dec}}$ }                    \\ \hline
					%\multicolumn{3}{|c|}{IDD Receiver}                                         & \multicolumn{2}{c|}{$C^{\text{Ini}}_{\text{ChE}} + C^{\text{Ini}}_{\text{Det}} + \left(I_{\text{IDD}}-1\right) C^{\text{Iter}}_{\text{Det}} + I_{\text{IDD}}C_{\text{Dec}}$ }                    \\ \hline
					\multicolumn{3}{|c|}{ICDD Receiver}                                        & \multicolumn{2}{c|}{$C^{\text{Ini}}_{\text{ChE}} + C^{\text{Ini}}_{\text{Det}} + \left(I_{\text{ICDD}}-1\right)\left(C^{\text{Iter}}_{\text{ChE}} + C^{\text{Iter}}_{\text{Det}}\right) + I_{\text{ICDD}}C_{\text{Dec}}$}                    \\ \hline
					\multicolumn{3}{|c|}{\textbf{JCDDNet-G}}                                & \multicolumn{2}{c|}{\makecell[c]{$L_{\text{JCDDNet-G}}\big(8N_t^3+8N_t^2+2N_t+16{N_r}{N_t^2}+16{N_t^2}T+16{N_r}{N_t}T+\left(8+2{d_c}\right)M2^{d_c-1}\big)$}}           \\ \hline
					\multicolumn{3}{|c|}{\textbf{JCDDNet-S}}                                & \multicolumn{2}{c|}{\makecell[c]{$L_{\text{JCDDNet-S}}\big(8{N_r}{N^2_t}+10{N_r}{N_t}+16{N_t^2}{T}+8{N_r}{N_t}{T}+\left(8+2{d_c}\right)M2^{d_c-1}\big)$}}           \\ \hline
				\end{tabular}%
			}
			\label{table1}
		\end{table*}
		
		\begin{table*}[!t]
			\centering
			
			\caption{Average CPU Time per Codeword of Different Receivers}
			{%
				\begin{tabular}{ccccccc}
					\toprule
					\multirow{3}{*}{	\textbf{Receivers}} & \multicolumn{6}{c}{	\textbf{Average CPU time (s)}}                                                          \\
					& \multicolumn{2}{c}{\begin{tabular}[c]{@{}c@{}}i.i.d. Gaussian \\ $N_r = 8$, $N_t = 4$\end{tabular}} & \multicolumn{2}{c}{\begin{tabular}[c]{@{}c@{}}Correlated Gaussian \\ $N_r = 8$, $N_t = 4$\end{tabular}} & \multicolumn{2}{c}{\begin{tabular}[c]{@{}c@{}}Sparse mmWave \\ $N_r = 64$, $N_t = 4$\end{tabular}} \\
					& QPSK (3 dB)& 16QAM  (10 dB)& QPSK (5 dB)& 16QAM    (13 dB) & QPSK  (12 dB)  & 16QAM  (22 dB)      \\
					\midrule
					MMSE-ICDD &      7.35e-1         &      3.43e-1         &      6.76e-1     &    2.38e-1          &     3.59e-1         &      4.58e-1       \\
					MMSE-ICDD (Selected) &      8.69e-1         &      3.86e-1         &      9.04e-1     &    2.92e-1          &     /         &      /       \\
					GAMP-ICDD &      5.38e-1        &     2.29e-1       &     1.30      &    1.44        &         1.96    &     2.20       \\
					MAP-ICDD &      6.14e-1        &      5.18       &    4.64e-1      &      4.49      &    2.85e-1          &    5.47        \\
					\textbf{JCDDNet-G} &    1.92e-1          &    4.25e-1            &      1.71e-1       &       3.03e-1     &           /   &         /     \\
					\textbf{JCDDNet-S}  &        /       &     /           &       /      &        /       &   1.15e-1         &  3.12e-1   \\
					\bottomrule        
				\end{tabular}%
			}
			\label{table2}
		\end{table*}		
		
		\vspace{-4mm}											\subsection{Complexity Analysis}
		The computational complexities of different receivers are analyzed. For clarity, both the concrete number of real-valued operations of different receivers and that of each module in traditional receivers are listed in Table \ref{table1}, where the number of layers of the proposed JCDD receivers, the number of turbo iterations, the number of ISTA iterations, the number of GAMP iterations, and the number of BP iterations, are denoted by $L_{\text{JCDDNet-G/S}}$, $I_{\text{IDD/ICDD}}$, $I_{\text{ISTA}}$,  $I_{\text{GAMP}}$, and $I_{\text{BP}}$, respectively. In addition, $C^\text{Ini}_\text{ChE}$ and $C^\text{Iter}_\text{ChE}$ are the complexities of the initial channel estimation based on the pilot symbols and the iterative channel estimation that utilizes the soft data symbols, $C^\text{Ini}_\text{Det}$ and $C^\text{Iter}_\text{Det}$ are the complexities of the initial data detection without  any prior information and the soft-in soft-output detection, while $C_\text{Dec}$ represents the decoding complexity with $d_c$ defined as the degree of check nodes. Note that compared with MMSE-ICDD, MMSE-ICDD (selcted) requires additional complexities  of ${\cal O}\left(\left(T_P+1\right)^3T_D\right)$ and ${\cal O}\left(N^3_r\left(T_P+1\right)^3T_D\right)$ for calculating the selection metrics corresponding to i.i.d. and correlated Gaussian channels, respectively, and a complexity of ${\cal O}\left(T_VT_D\right)$ for sorting.	
		As can be seen, the complexities of turbo receivers can be relatively high for a high modulation order or a large number of antennas, especially for MAP-ICDD of exponential complexity, while the complexities per layer of both the proposed JCDDNet-G and JCDDNet-S are almost linear with the codelength and independent of the modulation order, since $N_r, N_t \ll N$, $T$ and $M$ scale linearly with $N$, and ${d}_c$ is relatively small for LDPC codes. { Taking a Gaussian correlated MIMO channel with $N_r = 8$ and $N_t = 4$ for example, the LMMSE-based channel estimator requires 5.4e5 and 1.8e6 operations for $C^\text{Ini}_\text{ChE}$ and $C^\text{Iter}_\text{ChE}$, respectively, and the MAP-based detector requires 2.8e6 and 2.9e6 operations for $C^\text{Ini}_\text{Det}$ and $C^\text{Iter}_\text{Det}$ in the QPSK case, respectively, while the numbers of operations can be up to 3.8e8 and 4.0e8 in the 16QAM case. On the other hand, the proposed JCDD networks require only 1.1e5 operations per layer and a few layers can be sufficient for most receiving tasks as shown in Fig. \ref{fig4}(b).}
		\vspace{-2mm}
		
		To be more intuitive, the average central processing unit (CPU) time per codeword of the receivers with comparable performances is also listed in Table \ref{table2}, where the proposed JCDD networks are evaluated by loading the trained parameters into the corresponding ADMM algorithms implemented on the MATLAB platform. Note that for the QPSK cases, the required CPU time of MMSE-ICDD can be even longer than that of MAP-ICDD, since the complexity is dominated by the BP decoding for QPSK and MMSE-ICDD takes more turbo iterations, leading to a higher decoding complexity. We can also find that MMSE-ICDD (Selected) requires more time than MMSE-ICDD due to the introduced data selection steps. Moreover, although GAMP-ICDD runs relatively fast under i.i.d. Gaussian channels as expected, the required CPU time can be long under both correlated Gaussian channels and sparse mmWave channels, since it can barely benefit from the early-stopping criterion and takes almost the maximum number of turbo iterations in presence of channel correlation. On the other hand, the proposed JCDDNet-G and JCDDNet-S mainly involve linear multiplication/addition operations and are able to directly recover the bit variables within a few layers, which results in the shortest CPU time. For the 16QAM cases, we observe that MAP-ICDD needs an unbearably long CPU time due to the exponentially growing complexity with the modulation order. { Note that the less required time of MMSE-ICDD for 16QAM can be attributed to the early stopping mechanism and the fact that the number of symbols to detect is half of that in the QPSK case.} Additionally, the convergence rates of the JCDD networks become slower, which thus require a slightly longer CPU time than MMSE-ICDD. However, concerning the BLER performances, the proposed JCDD networks can always outperform MMSE-ICDD and  even approach the time-consuming MAP-ICDD, indicating a better tradeoff between the performance and the complexity. In fact, the proposed receiver design does not require the iterative soft information exchange between different modules, therefore avoiding the feedback delay. Moreover, most calculations involved in each layer of the proposed networks can be efficiently performed in parallel, making the proposed designs promising in the context of URLLC.
		\section{Conclusion}
		
		In this paper, by simultaneously incorporating the channel statistical information as well as the constraints imposed on pilots, data symbols, and LDPC codes, we proposed a unified MAP-based JCDD receiver framework to directly estimate the information bits. Specifically, two novel JCDD receivers for Gaussian MIMO channels and sparse mmWave MIMO channels were developed based on the ADMM algorithms, respectively, where closed-form solutions were derived for each ADMM iteration. Furthermore, the corresponding deep unfolding networks JCDDNet-G and JCDDNet-S were established to facilitate the optimization of the involved parameters. Simulation results validated that, both the proposed trainable receivers can always exhibit performance advantages over the MMSE-based turbo receivers and even achieve comparable performances to the MAP-based turbo receivers, while with a lower complexity per layer that is almost linear with the codelength and independent of the modulation order.
		
		\newcounter{TempEqCnt6}
		\setcounter{TempEqCnt6}{\value{equation}} 
		\setcounter{equation}{55} 	
		\begin{figure*}[ht] 
			%		\hrulefill
			\begin{small}
				\begin{align}		
					{\rm{tr}}\left({{{\bf{v}}_{{{\bf{b}}^{n-1}}}}{\bf{v}}_{{{\bf{b}}^{n-1}}}^H}{{\bf{X}}_{\bf{b}}^H}{{\bf{X}}_{\bf{b}}}\right) \mathop = \limits^{(a)}&{{\bf{v}}_{{{\bf{b}}^{n-1}}}^H}\left({{\left({{\bf S}_{\bf b}}{{\bf S}_{\bf b}^H}\right)}^T}\otimes{{\bf I}_{N_r}}\right){{\bf{v}}_{{{\bf{b}}^{n-1}}}} \nonumber 
					\mathop = \limits^{(b)}
					{\rm{tr}}\left({{\bf{V}}_{{{\bf{b}}^{n-1}}}}{{\bf S}_{\bf b}}{{\bf S}_{\bf b}^H}{{\bf{V}}_{{{\bf{b}}^{n-1}}}^H}\right)\nonumber\\
					\mathop= \limits^{(c)} &{{\rm{vec}}^H}\left({\bf S}_{\bf b}\right)\left( {{\bf{I}}_{T}} \otimes \left({{\bf{V}}_{{{\bf{b}}^{n-1}}}^H}{{\bf{V}}_{{{\bf{b}}^{n-1}}}}\right) \right){{\rm{vec}}}\left({\bf S}_{\bf b}\right)\nonumber\\
					\mathop \leq \limits^{(d)} 
					&{{\rm{vec}}^H}\left({\bf S}_{\bf b}\right)({\lambda ^{n-1}}{{\bf{I}}_{{N_t}T}}){\rm{vec}}\left({\bf S}_{\bf b}\right)\nonumber - 2{\mathop{\rm Re}\nolimits} \left\{ {{{\rm{vec}}^H}\left({\bf S}_{\bf b}\right)\left( {{\bf{I}}_{{N_t}T}}\otimes {{\left( {\lambda ^{n - 1}}{{\bf{I}}_{N_t}} - {{\bf{V}}_{{{\bf{b}}^{n-1}}}^H}{{\bf{V}}_{{{\bf{b}}^{n-1}}}} \right)}}   \right){\rm{vec}}\left({\bf S}_{{\bf b}^{n-1}}\right)} \right\}\nonumber \\ 
					&+ {{\rm{vec}}^H}\left({\bf S}_{{\bf b}^{n-1}}\right)\left( {{\bf{I}}_{{N_t}T}} \otimes{{\left( {\lambda ^{n - 1}}{{\bf{I}}_{N_t}} - {{\bf{V}}_{{{\bf{b}}^{n-1}}}^H}{{\bf{V}}_{{{\bf{b}}^{n-1}}}} \right)}}  \right){\rm{vec}}\left({\bf S}_{{\bf b}^{n-1}}\right)\nonumber\\		
					\mathop= \limits^{(e)} &{\lambda ^{n - 1}}\left\| f({\bf b}) \right\|_F^2 - 2{\mathop{\rm Re}\nolimits} \left\{ {{\rm{tr}}\left( {{f({\bf b})^H}}{\left( {\lambda ^{n - 1}}{{\bf{I}}_{{N_t}}} - {{\bf{V}}_{{{\bf{b}}^{n-1}}}^H}{{\bf{V}}_{{{\bf{b}}^{n-1}}}}  \right)f({{\bf b}^{n-1}})} \right)} \right\} + {\rm{constant}}, \label{eq22}
				\end{align}
			\end{small}
			\hrulefill
		\end{figure*} 
		\setcounter{equation}{\value{TempEqCnt6}}

		\begin{appendices}
			\newcounter{TempEqCnt5}
			\setcounter{TempEqCnt5}{\value{equation}} 
			\setcounter{equation}{52} 										
			
			{ 					
				\section{Gray Mapping Functions} \label{Ap_1}

				\begin{enumerate} 
					\item QPSK:
					\begin{equation} 
						\begin{small}
							\begin{aligned}
								f({{\bf{\tilde b}}^{t,k}}) = \frac{1}{{\sqrt 2 }}\left( {1 - 2{b_{2{i^{t,k}} - 1}}} \right) + \jmath\frac{1}{{\sqrt 2 }}\left( {1 - 2{b_{2{i^{t,k}}}}} \right),\label{eq9}
							\end{aligned}
						\end{small}
					\end{equation}
					\item 16QAM:
					\begin{equation}
						\begin{small}
							\begin{aligned} 
								f({{\bf{\tilde b}}^{t,k}}) = \frac{1}{{\sqrt {10} }} {\left( {1 - 2{b_{4{i^{t,k}} - 3}}} \right)\left( {1 +  2{b_{4{i^{t,k}} - 1}} } \right)} + \\
								\jmath \frac{1}{{\sqrt {10} }}{\left( {1 - 2{b_{4{i^{t,k}} - 2}}} \right)\left({1 + 2{b_{4{i^{t,k}}}}} \right)},\label{eq10}
							\end{aligned}
						\end{small}
					\end{equation}
				\end{enumerate} 
				where $\jmath \triangleq \sqrt{-1}$. }

			\setcounter{equation}{\value{TempEqCnt5}} 																								\section{Proof of Proposition 1} \label{Ap_2}
			To begin with, we define ${{\bf w}_{\bf b}} \triangleq {{\bf{X}}_{\bf b}^H}{\bf{y}} $ and ${{\bf R}_{\bf b}} \triangleq {{{\bf{X}}_{\bf b}^H}{{\bf{X}}_{\bf b}} + {\sigma ^2}{\bf{C}}_{\bf{g}}^{ - 1}}$, which gives
			\setcounter{equation}{54} 
			\begin{equation} 
				%\begin{small}
					\begin{aligned}		
						\Phi({\bf b})=&-{\bf{w}}_{\bf b}^H{\bf{R}}_{\bf b}^{- 1}{{\bf{w}}_{\bf b}}\\
						\mathop  \leq \limits^{(a)} &{\rm{tr}}\left({\bf{R}}_{{{\bf{b}}^{n-1}}}^{ - 1}{{\bf{w}}_{{{\bf{b}}^{n-1}}}}{\bf{w}}_{{{\bf{b}}^{n-1}}}^H{\bf{R}}_{{{\bf{b}}^{n-1}}}^{ - 1}{{\bf{R}}_{\bf{b}}}\right) \\&- 2{\mathop{\rm Re}\nolimits} \{ {\bf{w}}_{{{\bf{b}}^{n-1}}}^H{\bf{R}}_{{{\bf{b}}^{n-1}}}^{ - 1}{{\bf{w}}_{\bf{b}}}\}  - {\bf{w}}_{{\bf{b}}^{n-1}}^H{\bf{R}}_{{\bf{b}}^{n-1}}^{ - 1}{{\bf{w}}_{{\bf{b}}^{n-1}}} \\
						%&= 2{\mathop{\rm Re}\nolimits} \{ tr\left( {{\bf{q}}_{{{\bf{X}}^i}}^H{\bf{R}}_{{{\bf{X}}^i}}^{ - 1}\left(\frac{1}{{{\sigma ^2}}}{{\bf{X}}^H}{\bf{y}} + {\bf{\Sigma }}_{\bf{h}}^{ - 1}{\bf{\hat h}})} )\}
						%\\&\quad- tr( {{\bf{R}}_{{{\bf{X}}^i}}^{ - 1}{{\bf{q}}_{{{\bf{X}}^i}}}{\bf{q}}_{{{\bf{X}}^i}}^H{\bf{R}}_{{{\bf{X}}^i}}^{ - 1}(\frac{1}{{{\sigma ^2}}}{{\bf{X}}^H}{\bf{X}} + {\bf{\Sigma }}_{\bf{h}}^{ - 1})} ) + {\rm{constant}}\\
						\mathop  = \limits^{(b)}& {\rm{tr}}\left({{\bf{v}}_{{{\bf{b}}^{n-1}}}}{\bf{v}}_{{{\bf{b}}^{n-1}}}^H{{\bf{X}}_{\bf{b}}^H}{{\bf{X}}_{\bf{b}}}\right)-2{\rm{Re}}\{{\rm tr}\left( {\bf y}{\bf{v}}_{{{\bf{b}}^{n-1}}}^H{{\bf{X}}_{\bf{b}}^H}\right)\} \\ & + {\rm{constant}}, \label{eq21}
					\end{aligned}
				%\end{small}
			\end{equation}
			where (a) is obtained by exploiting the fact that $-{\bf{w}}_{\bf b}^H{\bf{R}}_{\bf b}^{- 1}{{\bf{w}}_{\bf b}}$ is jointly concave in ${{\bf w}_{\bf b}}$ and ${{\bf R}_{\bf b}}$ and therefore upperbounded by its first-order Taylor expansion, and (b) uses the definitions of ${{\bf w}_{\bf b}}$ and ${{\bf R}_{\bf b}}$ with ${{\bf{v}}_{{{\bf{b}}^{n-1}}}} \triangleq {\bf{R}}_{{{\bf{b}}^{n-1}}}^{ - 1}{{\bf{w}}_{{{\bf{b}}^{n-1}}}}={\left({{\bf{X}}_{{\bf b}^{n-1}}^H}{{\bf{X}}_{{\bf b}^{n-1}}} + {\sigma ^2}{\bf{C}}_{\bf{g}}^{-1}\right)^{ - 1}}{{\bf{X}}_{{\bf b}^{n-1}}^H}{\bf{y}}$. Furthermore, for the first term in (\ref{eq21}), we have (\ref{eq22}) shown at the top of the page, where ${\bf S}_{\bf b}=\left[ {{\bf{S}}_P,f({\bf b})} \right]$, (a) is acquired based on the definition of ${{\bf X}_{\bf b}} = {{\bf S}_{\bf b}^T}\otimes{{\bf I}_{N_r}}$ , ${\rm{tr}}\left({\bf A}{\bf B}\right) = {\rm{tr}}\left({\bf B}{\bf A}\right)$, and $\left({\bf A} \otimes {\bf B}\right)\left({\bf C} \otimes {\bf D}\right) = \left({\bf A}{\bf C}\right) \otimes \left({\bf B}{\bf D}\right)$, (b), (c), and (e) are derived using ${\rm{vec}}\left({\bf A}{\bf B}{\bf C}\right)=\left({{\bf C}^T} \otimes {\bf A}\right){\rm{vec}}\left({\bf B}\right)$ and ${\rm {tr}}\left({{\bf A}^H}{\bf B}\right) = {{{\rm{vec}}^H\left({\bf A}\right)}}{\rm{vec}}\left({\bf B}\right)$, ${{\bf{V}}_{{{\bf{b}}^{n-1}}}} = {\rm {unvec}}\left({{\bf{v}}_{{{\bf{b}}^{n-1}}}}\right) \in \mathbb{C}^{{N_r}\times{N_t}}$ with $\rm{unvec}(\cdot)$ denoting the inverse vectorization operation, and (d) is achieved by following \emph{Example 13} in \cite{MM8944280} with ${\lambda^{n-1}}{{\bf I}_{{N_t}}}\succeq {{\bf{V}}_{{{\bf{b}}^{n-1}}}^H}{{\bf{V}}_{{{\bf{b}}^{n-1}}}}$. Note that an appropriate choice is to set ${\lambda^{n-1}}$ to ${\lambda}_{\rm {max}}\left({{\bf{V}}_{{{\bf{b}}^{n-1}}}^H}{{\bf{V}}_{{{\bf{b}}^{n-1}}}}\right)$. Similarly, the second term in (\ref{eq21}) can be rewritten as
			\setcounter{equation}{56} 
			\begin{equation} 
				\begin{small}
					\begin{aligned}		
						&-2{\rm{Re}}\{{\rm tr}\left( {\bf y}{\bf{v}}_{{{\bf{b}}^{n-1}}}^H{{\bf{X}}_{\bf{b}}^H}\right)\} \\ = &-2{\rm{Re}}\{{\rm tr}\left({{f({\bf b})^H}}{{\bf{V}}_{{{\bf{b}}^{n-1}}}^H}{{\bf Y}_D}\right)\} + {\rm{constant}}.
						\label{eq23}
					\end{aligned}
				\end{small}
			\end{equation}
			Thus, we can arrive at a second majorization to $\Phi({\bf b})$ as
			\begin{equation} 
				\begin{small}
					\begin{aligned}		
						\Phi({\bf b})\leq & {\lambda ^{n - 1}}\left\| f({\bf{b}}) \right\|_F^2 - 2{\mathop{\rm Re}\nolimits} \left\{ {{\rm tr}\left( {{f({\bf{b}})}^H} {{\bf D}^{n-1}}\right)} \right\}  + {\rm{constant}},
						\label{eq24}
					\end{aligned}
				\end{small}
			\end{equation}
			where ${{\bf D}^{n-1}} = {\left( {\lambda ^{n - 1}}{{\bf{I}}_{{N_t}}} - {{\bf{V}}_{{{\bf{b}}^{n-1}}}^H}{{\bf{V}}_{{{\bf{b}}^{n-1}}}}  \right)f({{\bf b}^{n-1}})} + {{\bf{V}}_{{{\bf{b}}^{n-1}}}^H}{{\bf Y}_D}$. Hence, subproblem (\ref{eq20a}) boils down to the desired problem given in Proposition 1. 
			%the following surrogate problem:
			%			\begin{equation}
				%				% 		\begin{small}
					%				\begin{aligned}	
						%					{{\bf{b}}^{n}} = \mathop {\arg \min }\limits_{{\bf{b}} \in {{[0,1]}^N}} {\lambda ^{n - 1}}\left\| f({\bf{b}}) \right\|_F^2 - 2{\mathop{\rm Re}\nolimits} \left\{ {{\rm Re}\left( {{f({\bf{b}})}^H} {{\bf D}^{n-1}}\right)} \right\} \\ - \alpha \left\| {{\bf{b}} - {{\bf{0.5}}_N}} \right\|_2^2 + \frac{\mu }{2}{\left\| {\bf A}{\bf b}+ {{\bf{z}}^{n-1}} - {\bm \theta} + {{\bm \eta}^{n-1}} \right\|_2^2}, \label{eq59}
						%				\end{aligned}
					%				% 		\end{small}
				%			\end{equation}
			%			which is the desired result given in Proposition 1.
			
			\section{Detailed Expressions of ${{\beta}^{n-1}_{{i^{t,k,q}}}}$ and ${{\gamma}^{n-1}_{{i^{t,k,q}}}}$} \label{Ap_3}
			The expressions of ${{\beta}^{n-1}_{{i^{t,k,q}}}}$ and ${{\gamma}^{n-1}_{{i^{t,k,q}}}}$ in (\ref{eq28}) for QPSK and 16QAM are provided as follows:
			\begin{enumerate} 
				\item QPSK:
				\begin{subequations}
					\begin{equation}
						\begin{small}
							\begin{aligned}
								{\beta^{n - 1}_{{i^{t,k,1}}}} = 4{\lambda ^{n - 1}},{\gamma^{n - 1} _{{i^{t,k,1}}}} =  2\sqrt 2 {\mathop{\rm Re}\nolimits} \{ d_{kt}^{n - 1}\} - 2{\lambda ^{n - 1}},
							\end{aligned}
						\end{small}
					\end{equation}
					\begin{equation}
						\begin{small}
							\begin{aligned}
								{\beta^{n - 1} _{{i^{t,k,2}}}} = 4{\lambda ^{n - 1}},{\gamma^{n - 1} _{{i^{t,k,2}}}} =  2\sqrt 2 {\mathop{\rm Im}\nolimits} \{ d_{kt}^{n - 1}\} - 2{\lambda ^{n - 1}}.
							\end{aligned}
						\end{small}
					\end{equation}		\label{Ap1}
				\end{subequations}
				\item 16QAM:				
				\begin{subequations}
					\begin{equation}
						\begin{footnotesize}
							\begin{aligned}
								{\beta^{n - 1} _{{i^{t,k,1}}}} = &\frac{4}{5}{\lambda ^{n - 1}}{(1 + 2{b^{n - 1}_{{i^{t,k,3}}}})^2},\\ {\gamma^{n - 1} _{{i^{t,k,1}}}} =  &\frac{4}{{\sqrt {10} }}{\mathop{\rm Re}\nolimits} \{ d_{kt}^{n - 1}\} (1 + 2{b^{n - 1}_{{i^{t,k,3}}}}) - \frac{2}{5}{\lambda ^{n - 1}}{(1 + 2{b^{n - 1}_{{i^{t,k,3}}}})^2},
							\end{aligned}
						\end{footnotesize}
					\end{equation}				
					\begin{equation}
						\begin{footnotesize}
							\begin{aligned}					
								{\beta^{n - 1} _{{i^{t,k,2}}}} =& \frac{4}{5}{\lambda ^{n - 1}}{(1 + 2{b^{n - 1}_{{i^{t,k,4}}}})^2},\\ {\gamma^{n - 1} _{{i^{t,k,2}}}} =&  \frac{4}{{\sqrt {10} }}{\mathop{\rm Im}\nolimits} \{ d_{kt}^{n - 1}\} (1 + 2{b^{n - 1}_{{i^{t,k,4}}}}) - \frac{2}{5}{\lambda ^{n - 1}}{(1 + 2{b^{n - 1}_{{i^{t,k,4}}}})^2},
							\end{aligned}
						\end{footnotesize}
					\end{equation}				
					\begin{equation}
						\begin{footnotesize}
							\begin{aligned}
								{\beta^{n - 1} _{{i^{t,k,3}}}} =& \frac{4}{5}{\lambda ^{n - 1}}{(1 - 2{b^{n}_{{i^{t,k,1}}}})^2},\\ {\gamma^{n - 1} _{{i^{t,k,3}}}} =& - \frac{4}{{\sqrt {10} }}{\mathop{\rm Re}\nolimits} \{ d_{kt}^{n - 1}\} (1 - 2{b^{n}_{{i^{t,k,1}}}}) + \frac{2}{5}{\lambda ^{n - 1}}{(1 - 2{b^{n}_{{i^{t,k,1}}}})^2},
							\end{aligned}
						\end{footnotesize}
					\end{equation}				             
					\begin{equation}
						\begin{footnotesize}
							\begin{aligned}
								{\beta ^{n - 1}_{{i^{t,k,4}}}} =& \frac{4}{5}{\lambda ^{n - 1}}{(1 - 2{b^{n}_{{i^{t,k,2}}}})^2},\\ {\gamma^{n - 1} _{{i^{t,k,4}}}} =& -\frac{4}{{\sqrt {10} }}{\mathop{\rm Im}\nolimits} \{ d_{kt}^{n - 1}\} (1 - 2{b^{n}_{{i^{t,k,2}}}}) + \frac{2}{5}{\lambda ^{n - 1}}{(1 - 2{b^{n}_{{i^{t,k,2}}}})^2}.
							\end{aligned}
						\end{footnotesize}
					\end{equation}		\label{Ap2}
				\end{subequations}
			\end{enumerate}
		\end{appendices}	
		
	\end{sloppypar}
	
	\bibliography{JCDD.bib}

\end{document}